\documentclass[preprint,floatfix,preprintnumbers,superscriptaddress,showkeys,showpacs,nofootinbib,aps]{revtex4}

\usepackage{amssymb}
\usepackage{epsfig}
\usepackage{amssymb}
\usepackage{epsf}
\usepackage{epsfig}
\usepackage{color}
\usepackage{ifpdf}
\usepackage{longtable}

\newcommand{\nn}{\nonumber}
\newcommand{\ba}{\begin{eqnarray}}
\newcommand{\ea}{\end{eqnarray}}
\newcommand{\be}{\begin{equation}}
\newcommand{\ee}{\end{equation}}
\newcommand{\bd}{\begin{displaymath}}
\newcommand{\ed}{\end{displaymath}}

\newcommand{\Refs}[1]{Refs.~#1}

\newcommand{\mev}{\mathrm{MeV}}

\newcommand{\fm}{\mathrm{fm}}

\newcommand{\old}[1]{}

\newcommand{\plotangle}{270}

\def\desy{NIC, DESY, Platanenallee 6, D-15738 Zeuthen, Germany}
\def\Munster{Institut f\"ur Theoretische Physik, Universit\"at M\"unster,\\
Wilhelm-Klemm-Strasse 9, D-48149, Germany}

\begin{document}

\preprint{DESY 10-176}
\preprint{SFB/CPP-10-117}
\preprint{MS-TP-10-14}
\preprint{KEK-CP-242}
\preprint{JLAB-THY-10-1289}

\title{Resonance Parameters of the $\rho$-Meson from Lattice QCD}

%\begin{center}
%\includegraphics[width=100pt]{etmc}
%\end{center}\vspace{10pt}

\author{Xu Feng}
\altaffiliation{Current address: KEK.}
\affiliation{\desy}
\affiliation{\Munster}

\author{Karl Jansen}
\affiliation{\desy}

\author{Dru B.\ Renner}
\altaffiliation{Current address: Jefferson Lab.}
\affiliation{\desy}

\collaboration{ETMC Collaboration}
\noaffiliation

%\author[DESY,Munster]{Xu Feng\thanksref{KEK}},
%\author[DESY]{Karl Jansen},
%\author[DESY]{Dru B.\ Renner\thanksref{JLAB}}

%\address[DESY]{NIC, DESY, Platanenallee 6, D-15738 Zeuthen, Germany}
%\address[Munster]{Universit\"at M\"unster, Institut f\"ur Theoretische Physik,\\
%Wilhelm-Klemm-Strasse 9, D-48149, Germany}
%\thanks[KEK]{Current address: KEK.}
%\thanks[JLAB]{Current address: Jefferson Lab.}

\begin{abstract}
We perform a nonperturbative lattice calculation of
the P-wave pion-pion scattering phase in the $\rho$-meson decay channel
using two flavors of maximally twisted mass fermions at pion masses ranging
from 480 to 290 MeV. Making use of finite-size methods, we evaluate the
pion-pion scattering phase in the center-of-mass frame and two moving frames.
Applying an effective range formula,
we find a good description of our results for the scattering phase as 
a function of the energy covering the resonance region. This allows us 
to extract the $\rho$-meson mass and
decay width and to study their quark mass dependence.

\end{abstract}

\keywords{pion-pion scattering phase, $\rho$-meson resonance parameters, lattice QCD}
\pacs{14.40.Aq, 13.75.Lb, 12.38.Gc, 11.15.Ha}

\maketitle

\section{Introduction}

In experiments, many hadrons are observed as resonances that decay
via the strong interaction and have only a short life-time.
On the theoretical side, the direct determination of the resonance      
parameters from QCD is afflicted with many difficulties 
since the computation of resonance masses and decay widths is 
essentially a nonperturbative problem. An attractive way
to extract the resonance parameters nonperturbatively from first
principles is the use of lattice QCD. 
Among unstable hadrons,
the case of the $\rho$-meson is ideal for lattice studies
of a resonance for two reasons. First, in lattice calculations, the noise-to-signal ratio
in the computation of a meson mass
is proportional to $e^{(m_M-m_\pi)t}$, where $m_M$ is the meson mass under consideration,
$m_\pi$ is the pion mass and $t$ is a typical hadronic time scale. Since the $\rho$ is one of the lightest mesons,
the statistical error of its numerically computed mass can be well controlled.
Second, the principle decay channel (with a branching rate of 99.9\%) of the $\rho$-meson is to a pair of pions, which can be
treated on the lattice very precisely.
%As a result, a pion-pion scattering system provides an ideal laboratory for the study of rho resonance.

In the past, several lattice groups have undertaken efforts
to study the $\rho$-meson decay. A first attempt was made to
estimate the decay width from the $\rho\rightarrow\pi\pi$ transition
amplitude~\cite{Gottlieb:1983rh,Loft:1988sy,McNeile:2002fh,Jansen:2009hr}.
This method relies on two assumptions: first, the energy gap between the ground and the
first excited state (corresponding to $\rho$-meson and $\pi\pi$ states
with the same quantum numbers) is small. Second, it is assumed that the
hadron interaction is not large and the transition amplitude
$\langle\rho|\pi\pi\rangle$ satisfies
$\langle\rho|\pi\pi\rangle\ll \langle\rho|\rho\rangle^{1/2}\langle\pi\pi|\pi\pi\rangle^{1/2}$.
An alternative method, which does not rely on these assumptions, is to
extract the $\rho$-meson resonance parameters from the P-wave pion-pion scattering phase in the
isospin $I=1$ channel. The nonperturbative determination of the scattering phase is
possible by using finite-size methods, which were originally proposed by
L\"uscher in the center-of-mass frame
(CMF)~\cite{Luscher:1985dn,Luscher:1986pf,Luscher:1990ck,Luscher:1990ux,Luscher:1991cf}
and later extended to more general cases employing also a moving frame (MF)
by Rummukainen and Gottlieb~\cite{Rummukainen:1995vs}\footnote{In Ref.~\cite{Rummukainen:1995vs} the determination of scattering phase is 
accomplished at the level of quantum mechanics. For the field theorectic derivations we refer to Refs.~\cite{Kim:2005gf} and \cite{Christ:2005gi}.}. 
Making use of these finite-size methods,
two lattice studies~\cite{Aoki:2007rd,Gockeler:2008kc} 
have been carried out to compute the $\rho$-meson resonance
parameters.\footnote{We note that recently another 
two lattice studies~\cite{Aoki:2010hn,Frison:2010ws} were reported at Lattice 2010.} These calculations mainly
concentrated on the scattering phase at one or two energies for a single ensemble.
In this way, however, the scattering phase can be extracted at only a small number of energies
and it becomes difficult to map out the resonance region.

In this work, we study the $I=1$ pion-pion scattering system using three
Lorentz frames: the CMF, the first MF with total momentum ${\mathbf P}=(2\pi/L){\mathbf e}_3$ (MF1)
and the second MF with ${\mathbf P}=(2\pi/L)({\mathbf e}_1+{\mathbf e}_2)$ (MF2).
Here, the ${\mathbf e}_i$ denotes a unit vector in the spatial direction $i$ and
$L$ is the spatial extent of the lattice.
In each frame, we evaluate the P-wave scattering phase from the energy eigenvalues
of the ground state and the first excited state. Using three frames allows us
to obtain the scattering phase at six energies for each set of physical
parameters considered without the need to go to larger lattices.
Therefore, we think that our calculations have two advantages compared to
the earlier works mentioned above.
First,  extracting the resonance parameters from six energies allows us 
to obtain more accurate results. Second, some of the scattering phases are 
calculated at energies that lie in the range
$[m_\rho-\Gamma_\rho/2,m_\rho+\Gamma_\rho/2]$, allowing us to directly
map out the resonance region.

Our calculations are performed using the $N_f=2$ maximally twisted mass
fermion ensembles~\cite{Boucaud:2007uk,Boucaud:2008xu,Baron:2009wt}
from the European Twisted Mass Collaboration (ETMC) at a lattice spacing of
$a= 0.079~\mathrm{fm}$. The pion masses range from 290 up to 480 MeV,
ensuring that the physical kinematics for the $\rho$-meson decay, $m_\pi/m_\rho<0.5$,
is satisfied. The
computation of the $\rho$-meson resonance parameters at several values of the
pion mass allows us to obtain the pion mass dependence of the resonance
mass and decay width and hence to perform an extrapolation
to the physical point. The benefit of using twisted mass fermions is that
at maximal twist physical observables are automatically accurate to $O(a^2)$
in the lattice spacing, while the drawback is that isospin symmetry, although
again an $O(a^2)$ effect for the observables considered in this work, is broken
at nonzero values of the lattice spacing. As a result, for any value of
$a\ne 0$ the decay of $\rho^0$ to $\pi^0\pi^0$ is allowed, while
in the continuum limit isospin symmetry is restored and this decay is forbidden.
In this paper we present a first calculation to extract the $\rho$-meson
resonance parameters from three Lorentz frames and discuss the feasibility and accuracy
achievable using this setup.
Since here we use only one value of the lattice spacing,
we cannot test for the possible effects of isospin breaking.
We plan to come back to this issue in the future when we will analyze
gauge field ensembles obtained at finer values of the lattice
spacing.
As it will turn out, we are not able to match the high
experimental accuracy of the $\rho$-meson
resonance parameters with our lattice calculation. Still, we consider this
work an important conceptual study and the techniques used here
will be useful for other resonances such as the $\Delta$ baryon.
%Further details of checking isospin symmetry breaking effects at various
%lattice spacings will be shown in a forthcoming letter.

\section{Method}
\subsection{Scattering phase}
In an elastic scattering system, the relativistic Breit-Wigner form (RBWF) for the
scattering amplitude $a_l$ with a resonance at a center-of-mass (CM) energy $M_R$ 
and with a decay width $\Gamma_R$ is~\cite{Amsler:2008zzb}
\bd
a_l=\frac{-\sqrt{s}\Gamma_R(s)}{s-M_R^2+i\sqrt{s}\Gamma_R(s)}\;,\quad s=E_{CM}^2\;,
\ed
where $E_{CM}$ is the CM energy and $a_l$ is related
to the scattering phase of the $l^{th}$ partial wave, $\delta_l$, through $a_l=(e^{2i\delta_l}-1)/2i$.
The RBWF corresponding to $\delta_l$ is then
\be
\label{eq:BW}
\tan\delta_l=\frac{\sqrt{s}\Gamma_R(s)}{M_R^2-s}\;.
\ee

 The $\rho$-resonance has quantum numbers $I^G(J^{PC})=1^+(1^{--})$ and decays into two
pions in the P-wave. A description of the
scattering phase as a function of the $E_{CM}$ is provided by the effective
range formula (ERF)~\cite{Brown:1968zz}
    \be
    \label{eq:effective_range_formula}
    \tan{\delta_1}=\frac{g^2_{\rho\pi\pi}}{6\pi}\frac{p^3}{E_{CM}(m_\rho^2-E_{CM}^2)}
    \;,\quad p=\sqrt{E_{CM}^2/4-m_\pi^2}\;,
    \ee
which fits the experimental data well. In Eq.~(\ref{eq:effective_range_formula}) $\delta_1$
is the P-wave pion-pion
scattering phase, $g_{\rho\pi\pi}$ is the effective $\rho\rightarrow\pi\pi$ coupling
constant and $m_\rho$ is the $\rho$-meson mass. We remark already at this point that we will
use the ERF also for our lattice calculations to fit the scattering phase,
even when using pion masses that are larger than the physical one.
Comparing Eqs.~(\ref{eq:BW}) and (\ref{eq:effective_range_formula}), we find that the
ERF is a particular case of the RBWF if the parameters $M_R$ and $\Gamma_R(s)$ are chosen such that
 \bd
 M_R=m_\rho\;,\quad\Gamma_R(s)=\frac{g^2_{\rho\pi\pi}}{6\pi}\frac{p^3}{s}\;.
 \ed
The rho decay width $\Gamma_\rho$ can then be computed in the following way,
 \be
 \label{eq:decay_width}
 \Gamma_\rho=\Gamma_R(s)\Bigg|_{s=m_\rho^2}=\frac{g^2_{\rho\pi\pi}}{6\pi}\frac{p^3_\rho}{m_\rho^2}\;,\quad p_\rho=\sqrt{m_\rho^2/4-m_\pi^2}\;.
 \ee
 Thus Eqs.~(\ref{eq:effective_range_formula}) and (\ref{eq:decay_width}) allow us to
extract $m_\rho$ and $\Gamma_\rho$ by studying the dependence of the pion-pion scattering
phase $\delta_1$ on $E_{CM}$.

 \subsection{Finite-size methods}
 \subsubsection{Center-of-mass frame}
 A direct calculation of the phase shift from lattice QCD is possible by using
a finite-size method established by
L\"uscher~\cite{Luscher:1985dn,Luscher:1986pf,Luscher:1990ck,Luscher:1990ux,Luscher:1991cf}.
In this method, the phase shift is obtained from the energy eigenvalues of a two-pion
system enclosed in a cubic box with spatial size $L$.

 In the CMF the possible energy eigenvalues for two noninteracting pions are given by
 \bd
 \bar{E}=2\sqrt{m_\pi^2+\bar{p}^2}\;,\quad\bar{p}=|\bar{\mathbf p}|\;,\quad \bar{\mathbf p}=(2\pi/L){\mathbf n}\;,\quad {\mathbf n}\in \mathbb{Z}^3\;.
 \ed
 In the interacting case the energy eigenvalues are shifted,
 \bd
 E=2\sqrt{m_\pi^2+p^2}\;,\quad p=(2\pi/L)q\;,%\quad q\neq |{\mathbf n}|\;.
 \ed
where $q$ is no longer constrained to originate from a quantized momentum mode.
Because of the presence of the interaction, the energy eigenvalues deviate from
those in the noninteracting case.
It is exactly this
deviation that contains the information of the underlying strong interaction
and thus can be used to determine the scattering phase, as outlined next.

In this paper, we concentrate on the energy eigenstates with energies $E$ in the
elastic region $2m_\pi<E<4m_\pi$ with the two-pion system having the same
quantum numbers as the $\rho$-meson. In the CMF these states transform 
as a vector (more specifically the irreducible representation $\Gamma=T_1^-$)
under the cubic group $O_h$. The corresponding finite-size formula connecting
the energy $E$ to the scattering phase $\delta_1$ is given by
\cite{Luscher:1991cf}
 \be
 \label{eq:CMF}
 \tan\delta_1(E)=\frac{\pi^{3/2}q}{\mathcal{Z}_{00}(1;q^2)}\;,\quad \textmd{for }\;\Gamma=T_1^-\;,
 \ee
 with the zeta function defined through
 \bd
 \mathcal{Z}_{00}(s;q^2)=\frac{1}{\sqrt{4\pi}}\sum_{{\mathbf n}\in\mathbb{Z}^3}\left(|{\mathbf n}|^2-q^2\right)^{-s}\;.
 \ed

 \subsubsection{Moving frame}
Using a MF with total momentum ${\mathbf P}=(2\pi/L){\mathbf d}$,
${\mathbf d}\in\mathbb{Z}^3$, the energy eigenvalues in the
noninteracting case are given by
 \bd
 \bar{E}=\sqrt{m_\pi^2+\bar{p}_1^2}+\sqrt{m_\pi^2+\bar{p}_2^2}\;,
 \ed
 where $\bar{p}_i=|\bar{{\mathbf p}}_i|$ and $\bar{{\mathbf p}}_i$ denote
the three-momenta of the pions, which satisfy the relations
 \be
 \label{eq:sum_p}
 \bar{{\mathbf p}}_i=(2\pi/L){\mathbf n}_i\;,\quad {\mathbf n}_i\in \mathbb{Z}^3\;,\quad\bar{{\mathbf p}}_1+\bar{{\mathbf p}}_2={\mathbf P}\;.
 \ee
 In the MF, the center-of-mass is moving with a velocity of ${\mathbf v}={\mathbf P}/\bar{E}$. 
 Using the standard Lorentz transformation with a boost factor $\gamma=1/\sqrt{1-{\mathbf v}^2}$, 
 the $\bar{E}_{CM}$ can be obtained as
 \bd
 \bar{E}_{CM}=\gamma^{-1}\bar{E}=2\sqrt{m_\pi^2+\bar{p}^{*2}}\;,
 \ed
 with CM momenta
 \be
 \label{eq:sub_p}
 \bar{p}^*=|\bar{\mathbf p}^*|\;,\quad
 \bar{\mathbf p}^*=\bar{\mathbf p}^*_1=-\bar{\mathbf p}^*_2=\frac{1}{2}\vec{\gamma}^{-1}(\bar{{\mathbf p}}_1-\bar{{\mathbf p}}_2)\;.
 \ee
Here, we use the notation
 \bd
% \vec{\gamma}{\mathbf p}=\gamma{\mathbf p}_{\parallel}+{\mathbf p}_{\perp}\;,\quad
 \vec{\gamma}^{-1}{\mathbf p}=\gamma^{-1}{\mathbf p}_{\parallel}+{\mathbf p}_{\perp}\;,\quad
 {\mathbf p}_{\parallel}=\frac{{\mathbf p}\cdot{\mathbf v}}{|\mathbf v|^2}{\mathbf v}\;,\quad
 {\mathbf p}_{\perp}={\mathbf p}-{\mathbf p}_{\parallel}\;.
 \ed
From inspecting Eqs.~(\ref{eq:sum_p}) and (\ref{eq:sub_p}) it can be seen
that the $\bar{\mathbf p}^*$ are quantized to the values
 \be
 \label{eq:set_of_momentum}
 \bar{\mathbf p}^*=(2\pi/L){\mathbf n}\;,\quad {\mathbf n}\in P_{\mathbf d}=\left\{{\mathbf n}\left|\;{\mathbf n}=\vec{\gamma}^{-1}({\mathbf m}+{\mathbf d}/2)\;,\;\;\textmd{for}\;{\mathbf m}\in\mathbb{Z}^3\right.\right\}\;.
 \ee
 In the interacting case, the $E_{CM}$ is given by
 \be
 \label{eq:continuum_dis_rel}
 E_{CM}=2\sqrt{m_\pi^2+p^{*2}}\;,\quad p^*=(2\pi/L)q\;. %,\quad q\neq|{\mathbf n}|\;.
 \ee
 From the energy shift between the noninteracting and the interacting situation,
$E_{CM}-\bar{E}_{CM}$ (or equivalently $q^2-|{\mathbf n}|^2$), one can compute
the pion-pion scattering phase.

 In the MF1 (${\mathbf d}={\mathbf e}_3$), the energy eigenstates transform under the tetragonal group
 $D_{4h}$. The irreducible representations $A_2^-$ and $E^-$ are relevant for
the pion-pion scattering states $|\pi\pi,l=1\rangle$ 
in infinite volume with angular momentum $l=1$. In this work, we calculate the
energies associated with the $A_2^-$ sector. The formula converting the $E_{CM}$ in a
finite volume to the scattering phase in the infinite volume is given by
Gottlieb and Rummukainen \cite{Rummukainen:1995vs} as
 \be
 \label{eq:MF1}
 \tan\delta_1(E_{CM})=\frac{\gamma\pi^{3/2}q}{\mathcal{Z}_{00}^{\mathbf d}(1;q^2)+(2q^{-2}/\sqrt{5})\mathcal{Z}_{20}^{\mathbf d}(1;q^2)}\;,\quad \textmd{for }\;\Gamma=A_2^-\;,
 \ee
 with the modified zeta function
 \bd
 \mathcal{Z}_{lm}^{\mathbf d}(s;q^2)=\sum_{{\mathbf n}\in P_{\mathbf d}}\frac{\mathcal{Y}_{lm}^*({\mathbf n})}{\left(|{\mathbf n}|^2-q^2\right)^{s}}
 \ed
and
 \bd
 \mathcal{Y}_{lm}(\mathbf r)\equiv r^lY_{l,m}(\Omega_r)\;,\quad \mathcal{Y}_{l\bar{m}}(\mathbf r)\equiv r^lY_{l,-m}(\Omega_r)\;,
 \ed
 where $\Omega_r$ represents the solid angle parameters ($\theta,\phi$) of $\mathbf r$ in spherical coordinates and the $Y_{l,m}$ are the usual spherical harmonic functions.

In order to obtain more energies in the resonance region, we developed a
second moving frame (MF2) with ${\mathbf d}={\mathbf e}_1+{\mathbf e}_2$. The corresponding energy eigenstates transform
under the orthorhombic group $D_{2h}$. The irreducible representations
$B_1^-$, $B_2^-$ and $B_3^-$ occur for the $|\pi\pi,l=1\rangle$ states
in infinite volume. Here we focus on the $B_1^-$ sector. Our derivation of the
corresponding finite-size formula for the MF2 results in
 \ba
 \label{eq:MF2}
 &&\tan\delta_1(E_{CM})=\frac{\gamma\pi^{3/2}q}{\mathcal{Z}_{00}^{\mathbf d}-(q^{-2}/\sqrt{5})\mathcal{Z}_{20}^{\mathbf d}+i(\sqrt{3}q^{-2}/\sqrt{10})\left(\mathcal{Z}_{22}^{\mathbf d}-\mathcal{Z}_{2\bar{2}}^{\mathbf d}\right)}\;,\nn\\
 &&\hspace{9.5cm}\textmd{for }\;\Gamma=B_1^-\;.
 \ea
For more details, we refer the reader to Ref.~\cite{Feng:2011ah}.

 For brevity, we represent $\mathcal{Z}_{lm}^{\mathbf d}(1;q^2)$ with the short 
 notation $\mathcal{Z}_{lm}^{\mathbf d}$ in Eq.~(\ref{eq:MF2}).
 Using Eqs.~(\ref{eq:CMF}), (\ref{eq:MF1}) and (\ref{eq:MF2}) we can then convert a
finite-volume determination of the $E_{CM}$ into a calculation of
the P-wave scattering phase $\delta_1$.
This is, of course, exactly the situation we are confronted with in a
lattice calculation as performed here.

\subsection{Correlation matrix}

 In the CMF, the value of the $E_{CM}$ is directly given
 by the discrete energy eigenvalue $E$ extracted from the large time behavior of the
 corresponding correlation
 function. In the MF,
 $E_{CM}$ is related to $E$ through the Lorentz transformation
  \be
  \label{eq:Lorentz_cont}
  E_{CM}^2=E^2-{\mathbf P}^2\;.
  \ee

 In order to calculate the energy eigenvalues $E$, we construct a
 $2\times2$ correlation function matrix through
     \be
     \label{eq:correltaion_matrix}
     C_{2\times2}(t)=\left(
     \begin{array}{cc}
     \left\langle \left(\pi\pi\right)(t)\;\left(\pi\pi\right)^\dagger(0)\right\rangle
     & \left\langle \left(\pi\pi\right)(t)\;\rho^\dagger(0)\right\rangle\\
     \left\langle\rho(t)\;(\pi\pi)^\dagger(0)\right\rangle
     & \left\langle\rho(t)\;\rho^\dagger(0)\right\rangle \\
     \end{array}
     \right)\;.
     \ee

 \subsubsection{$\pi\pi$ sector}
 The $\pi\pi$
 correlation function is constructed with the interpolating operators defined through
 \ba
 \label{eq:operator_construct}
 (\pi\pi)(t)&=&\frac{d_\Gamma}{N_G}\sum_{\hat{R}\in G}\chi_\Gamma(\hat{R})
 \left(\pi^+({\mathbf P}/2+\hat{R}{\mathbf p},t)\pi^-({\mathbf P}/2-\hat{R}{\mathbf p},t)\right.\nn\\
 &&\hspace{2.5cm}\left.-\pi^-({\mathbf P}/2+\hat{R}{\mathbf p},t)\pi^+({\mathbf P}/2-\hat{R}{\mathbf p},t)\right)\;,
 \ea
 with the momenta on the lattice $\mathbf P$ and
 ${\mathbf p}$ taking discrete values
 \bd
 \label{eq:constraint_mom}
 {\mathbf P}=(2\pi/L){\mathbf d}\;,\quad{\mathbf p}={\mathbf P}/2+(2\pi/L){\mathbf m}\;,\quad
 \textmd{for}\;\;{\mathbf d},{\mathbf m}\in\mathbb{Z}^3\;.
 \ed

Let us explain the notation we have used in Eq.~(\ref{eq:operator_construct}).
The pion interpolating operator
$\pi^\pm({\mathbf q},t)$
is defined through
 \bd
 \pi^a({\mathbf q},t)=\frac{1}{L^{3/2}}\sum_{\mathbf x}e^{-i{\mathbf q}\cdot{\mathbf x}}
 \left(\bar{\psi}\gamma_5\frac{\tau^a}{2}\psi\right)({\mathbf x},t)\;,\quad a=\pm,0\;,
 \ed
where $\tau^a$ denote the isospin Pauli matrices and $\psi$ the two-flavor quark fields.
We also introduce
the symmetry group $G$ as the set of all lattice rotations and reflections $\hat{R}$, under which the set of $P_{\mathbf d}$ 
defined by Eq.~(\ref{eq:set_of_momentum}) is invariant 
 \be
 \label{eq:constraint2}
 G=\left\{\hat{R}\left|\;\hat{R}{\mathbf n}\in P_{\mathbf d}\;,\;\;\forall\;{\mathbf n}\in P_{\mathbf d}\right.\right\}.
 \ee

In the
 CMF, MF1 and MF2, $G$ is given by the cubic groups $O_h$, $D_{4h}$ and $D_{2h}$, respectively.
 $\Gamma$ is the irreducible representation of the group $G$, $d_\Gamma$ is the dimension of $\Gamma$ and
$\chi_\Gamma(\hat{R})$ is the character of $\Gamma$.
 The average over all the operations $\hat{R}$ in the group $G$ weighted by the coefficient
 $\chi_\Gamma(\hat{R})$ projects out the scattering states
that belong to the $\Gamma$ representation.
Finally,
$N_G=\sum_{\hat{R}\in G} 1$. 

Given the momenta $\{{\mathbf P},{\mathbf p}\}$ and the representation $\Gamma$, one
can construct the interpolating operators $(\pi\pi)(t)$ using Eq.~(\ref{eq:operator_construct}).
Here we set $\Gamma$ to be $T_1^-$, $A_2^-$ and $B_1^-$ for the CMF,
MF1 and MF2, respectively, so that the energy eigenstates $|\pi\pi,\Gamma\rangle$ in finite volume
will approximate the P-wave scattering states $|\pi\pi,l=1\rangle$ in infinite volume 
if one ignores states with higher angular momentum. 
%Besides, we 
%choose the discrete value of $\mathbf m$ in Eq.~(\ref{eq:constraint_mom}) to be 
%${\mathbf e}_3$
%for the CMF and ${\mathbf 0}$ for the two MFs. With these setup, we construct
% the operators used in our calculation.
 In the CMF, the interpolating operator is given by
 \bd
 (\pi\pi)(t)=\pi^+\left(\frac{2\pi}{L}{\mathbf e}_3,t\right)\pi^-\left(-\frac{2\pi}{L}{\mathbf e}_3,t\right)
 -\pi^+\left(-\frac{2\pi}{L}{\mathbf e}_3,t\right)\pi^-\left(\frac{2\pi}{L}{\mathbf e}_3,t\right)\;.
 \ed
In the two MFs,
the operators are given in a unified form through
 \bd
 (\pi\pi)(t)=\pi^+\left({\mathbf P},t\right)\pi^-\left({\mathbf 0},t\right)
 -\pi^+\left({\mathbf 0},t\right)\pi^-\left({\mathbf P},t\right)\;,
 \ed
with ${\mathbf P}$ again the total threa-momentum of the scattering system. We can use
these
 operators to measure the energy eigenvalues $E$ from the corresponding
correlation functions, convert $E$ into $E_{CM}$ by applying Eq.~(\ref{eq:Lorentz_cont}) and then extract the
P-wave scattering phase $\delta_1$ using the finite-size formulae listed above.

 \subsubsection{$\rho$ sector}
 The interpolating operator for the neutral $\rho$-meson is constructed
 through a local vector current,
 \bd
 \label{eq:definition_rho}
 \rho(t)=\rho^0({\mathbf P},t)=\frac{1}{L^{3/2}}\sum_{{\mathbf x}}e^{-i{\mathbf P}\cdot{\mathbf x}}
 \left(\bar{\psi}({\mathbf a}\cdot{\mathbf \gamma})\frac{\tau^0}{2}\psi\right)({\mathbf x},t)\;,\quad
 {\mathbf a}\cdot{\mathbf \gamma}=\sum_{i=1}^{3}a_i\gamma_i\;,
 \ed
 where ${\mathbf a}$ indicates the polarization of the
vector current.
%To guarantee total momentum conservation,
% the summation is taken over spatial position ${\mathbf x}$ with a
%factor $e^{-i{\mathbf P}\cdot{\mathbf x}}$,
% which constrains the considered states to those with momentum ${\mathbf P}$.
 The direction of ${\mathbf a}$ is taken to be
 parallel to ${\mathbf e}_3$ in the CMF,
 ${\mathbf e}_3$ in the MF1 and
 ${\mathbf e}_1+{\mathbf e}_2$ in the MF2, respectively.
This choice allows us to obtain a good signal-to-noise ratio for the
off-diagonal matrix element $\left\langle\rho(t)\;(\pi\pi)^\dagger(0)\right\rangle$ in
Eq.~(\ref{eq:correltaion_matrix}).

\subsection{Extraction of energies}

By computing the matrix of correlation functions in Eq.~(\ref{eq:correltaion_matrix}),
we are able to isolate the ground state
 and first excited state in a clean way. This is of particular importance in the
resonance region, where the
 avoided level crossing occurs and the first excited state is potentially
 close to the ground state. Such a situation renders the
 extraction of the ground state energy difficult when only a
 single exponential fit ansatz is used.
Since we cannot predict a priori whether our energy levels will be
close to the resonance region, we find it necessary to always use
the correlation matrix to analyze our results.
To extract the energy eigenstates, we follow the variational
method~\cite{Luscher:1990ck} and construct a ratio of correlation function matrices as
 \bd
 \label{eq:variational_method}
 R(t,t_R)=C_{2\times2}(t)C^{-1}_{2\times2}(t_R)\;,\quad \textmd{for}\;t>t_R\;,
 \ed
 where $t_R$, the
 reference time slice, is assumed to be large enough such that the
 contributions to the matrix $R(t,t_R)$ from the excited states $|n\rangle$ with $n>2$ can be ignored.

 The two eigenvalues $R_n(t,t_R)$ ($n=1,2$) of the matrix
 $R(t,t_R)$ behave as
 \be
  \label{eq:correlator_R_n}
 R_n(t,t_R)\rightarrow A_n\cosh\left(-E_n (t-T/2)\right)\;,
 \ee
 where we assume that $t$ is large enough ($t>t_R\gg0$) to neglect excited states but still far enough from the boundaries ($t\ll T/2$) to
 ignore the unwanted thermal contributions as discussed in the case
for the pion scattering length using twisted mass fermions in Ref.~\cite{Feng:2009ij}.

 \section{Lattice calculation}
 \subsection{Ensemble information}

The results presented here
are from a sequence of ensembles with a lattice spacing of
$a=0.079~\fm$.  The pion masses range from $m_\pi=480$ to
$290~\mev$. At all pion masses the physical kinematics of
$m_\pi/m_\rho<0.5$ is satisfied, such that it is
physically possible for the $\rho$-meson to decay into two pions.
In our analysis we use two lattice sizes.
The first corresponds to
$L=1.9~\fm$ with pion masses of $m_\pi=480~\mev$ and $m_\pi=420~\mev$,
i.e.\ ensembles $A_1$ and $A_2$ in Table~\ref{tab:ensemble1}.
The second uses
$L=2.5~\fm$ with pion masses of $m_\pi=330~\mev$ and $m_\pi=290~\mev$,
i.e.\ ensembles $A_3$ and $A_4$ in Table~\ref{tab:ensemble1}.
Additional information about the ensembles used is given 
in Table~\ref{tab:ensemble1} and in
\Refs{\cite{Boucaud:2007uk,Boucaud:2008xu,Baron:2009wt}}.

\begin{table}[tb]
\begin{center}
\begin{tabular}{|c|c|c|c|c|c|c|}
\hline Ensemble & $\beta$ & $a\mu$ & $L/a$ & $m_\pi$ &$m_{\pi}/m_\rho$ & $N$ \\
% \hline 3.90  & 0.0100 & $24$ & 520 & 2.77(2) & 479 & 7.23(59)(41) & -0.297(20)(16) \\
 \hline $A_1$ & 3.90  & 0.0085 & $24$ & 480 & 0.43 & 176  \\
 \hline $A_2$ & 3.90  & 0.0064 & $24$ & 420 & 0.40 & 278 \\
 \hline $A_3$ & 3.90  & 0.0040 & $32$ & 330 & 0.32 & 124 \\
 \hline $A_4$ & 3.90  & 0.0030 & $32$ & 290 & 0.30 & 129 \\
% \hline $B_1$ & 4.05  & 0.0030 & $32$ & 320 & 0.32(2) & 138 \\
 \hline
\end{tabular}
\end{center}
% This is a hack that we should fix somehow.
\vspace{10pt} \caption{Ensembles used in this work. We give the ensemble name $A_i$,
the inverse bare coupling $\beta=6/g_0^2$,
the bare quark mass $a\mu$, the lattice size $L/a$ and the value of $m_\pi$ in units of MeV.
We also list
the ratio $m_\pi/m_\rho$ and the number $N$ of configurations used.}
\label{tab:ensemble1}
\end{table}

\subsection{Sources}

To calculate the $\pi\pi$ correlation functions, we employ a
stochastic method using $Z_4$ noise sources $\xi^i_{t_s}(\mathbf x)$ 
that are restricted to each three-dimensional time-slice with 
time $t_s$. The sources $\xi^i_{t_s}(\mathbf x)$ are also diluted in the color and spin indices, which are suppressed for simplicity. 
%The sources are restricted to spin, color and 3-dimensional spatial space, where the spin and color indices are suppressed. 
The index $i$ runs from $1$ to $N_s$, the  
number of stochastic noise sources.  In this work we are able to achieve sufficient accuracy with just $N_s=1$ samples. 
%For a given momentum mode $\mathbf q$ 
%we construct the source with momentum as 
%$\xi^i_{t_s}({\mathbf q},{\mathbf x})=e^{i{\mathbf q}\cdot{\mathbf x}}\xi^i_{t_s}(\mathbf x)$.
Using the one-end trick~\cite{Foster:1998vw}, we need to introduce two stochastic noise sources, 
$e^{i{\mathbf q}\cdot{\mathbf x}}\xi^i_{t_s}(\mathbf x)$ and $\xi^i_{t_s}(\mathbf x)$, for each
factor of $\pi^{\pm}({\mathbf q},t_s)$ in the correlation function.  For the correlation
functions in the MFs, we must account for two momentum modes ($e^{i{\mathbf q}\cdot{\mathbf x}}\xi^i_{t_s}(\mathbf x)$, ${\mathbf q}={\mathbf 0}$ and ${\mathbf q}={\mathbf P}$).
In the CMF there are three required momentum modes 
($e^{i{\mathbf q}\cdot{\mathbf x}}\xi^i_{t_s}(\mathbf x)$, ${\mathbf q}={\mathbf 0}$, ${\mathbf q}=(2\pi/L){\mathbf e}_3$ and ${\mathbf q}=-(2\pi/L){\mathbf e}_3$).
Since we place the source on
all the time slices $t_s=0, \ldots, T\!-\!1$, we therefore perform $T$
inversions for each configuration and each momentum mode.
Note that the time extent of our lattices is chosen to be always twice
the spatial extent.
The correlator $C_{11}(t)$ is then calculated
through
\bd C_{11}(t)=\left\langle
\left(\pi\pi\right)(t)\left(\pi\pi\right)^\dagger(0)\right\rangle=
\frac{1}{T}\sum_{t_s}\left\langle
\left(\pi\pi\right)(t+t_s)\left(\pi\pi\right)^\dagger(t_s)\right\rangle\;.
\ed
The rather large effort to generate propagators on all the
time slices allows us to
obtain the
correlators with high precision, which is important to extract
the desired energies reliably.

In the calculation of the off-diagonal correlator, $C_{21}(t)$, the
contraction of the quark fields leads to a three-point diagram.
Since in this three-point diagram the two-pion fields are located at the same
source time slice $t_s$, we use the sequential propagator method to
construct the correlator. We calculate $C_{21}(t)$ through
\bd
C_{21}(t)=\left\langle\rho(t)(\pi\pi)^\dagger(0)\right\rangle=
\frac{1}{T}\sum_{t_s}\left\langle\rho(t+t_s)(\pi\pi)^\dagger(t_s)\right\rangle\;,
\ed
and again average the correlator over all time slices $t_s$. 
For the second off-diagonal correlator $C_{12}(t)$, the two-pion
fields are placed at the sink time slice $t+t_s$, which would render
the computation of $C_{12}(t)$ more difficult. However, using the 
relation $C_{12}(t)=C_{21}^\ast(t)$, we get the off-diagonal matrix
element $C_{12}$ for free.
%which would require $T$ sequential propagators for each source time
%slice.
%However, using the relation $C_{12}(t)=C_{21}^\ast(t)$, we get the
%off-diagonal matrix element $C_{12}$ for free,
%thus saving computer time.

For the $\rho$-correlator, $C_{22}(t)$, we have performed a comparison
between the $Z_4$ stochastic source method and the point source
method and found that the
required computational effort to achieve a given signal-to-noise ratio
is comparable.
Historically, we started our work with the calculation of the
hadronic vacuum polarization tensor~\cite{Renner:2009by}. Since in that
work we generated point source propagators for the ensembles
listed in Table~{\ref{tab:ensemble1}}, we just use the available
propagators to construct the $\rho$-correlator
\bd
C_{22}(t)=\left\langle\rho^\dagger(t+t_s)\rho(t_s)\right\rangle\;,
\ed
where now the source time slices, $t_s$, are chosen randomly to
reduce the autocorrelation between consecutive gauge field configurations.

Because of the isospin symmetry breaking effects at nonzero lattice
spacing in our maximally twisted mass setup, the disconnected
diagram for the neutral $\rho$-meson does not vanish.
To address the disconnected contribution
to the neutral $\rho$-meson, we need to generate,
in principle,
additional all-to-all propagators. However, the disconnected diagram
correction has been studied in Ref.~\cite{Michael:2007vn} and has been found
to be negligibly small, and hence we neglect it also here in the
computation of the neutral $\rho$-correlator. For the same reason, we
neglect the disconnected diagrams for the off-diagonal entries, 
where these contributions originate solely from the neutral $\rho$-operator.
In the calculation of the correlator $\langle(\pi\pi)(\pi\pi)^\dagger\rangle$,
we are able to address these
disconnected pieces, since we put stochastic sources on all the time slices. 
%However,
%in the calculation of the correlator 
%$\langle(\pi\pi)(\pi\pi)^\dagger\rangle$, there are disconnected contributions that do not vanish in the continuum limit, but we are able to address these
%disconnected pieces since we put stochastic sources on all the time slices. 
We find that the disconnected diagram makes an apparently small
contribution to the correlator but adds a significant amount of noise, which 
would destroy the signal for the connected piece. Therefore we drop it
from the $\pi\pi$ sector.
To be clear, neglecting these disconnected contributions is not a genuine approximation but
is simply ignoring lattice artifacts that would vanish in the
continuum limit anyway.
%Remembering that this work is an exploratory study, addressing
%conceptual issues of determining resonance parameters from lattice calculations,
%we think that it is fully justified to neglect these contributions
%to the correlation function matrix.

\section{Results}
\subsection{Energy eigenvalues}

\begin{figure}
\begin{center}
%\begin{tabular}{cc}
\includegraphics[width=150pt,angle=\plotangle]{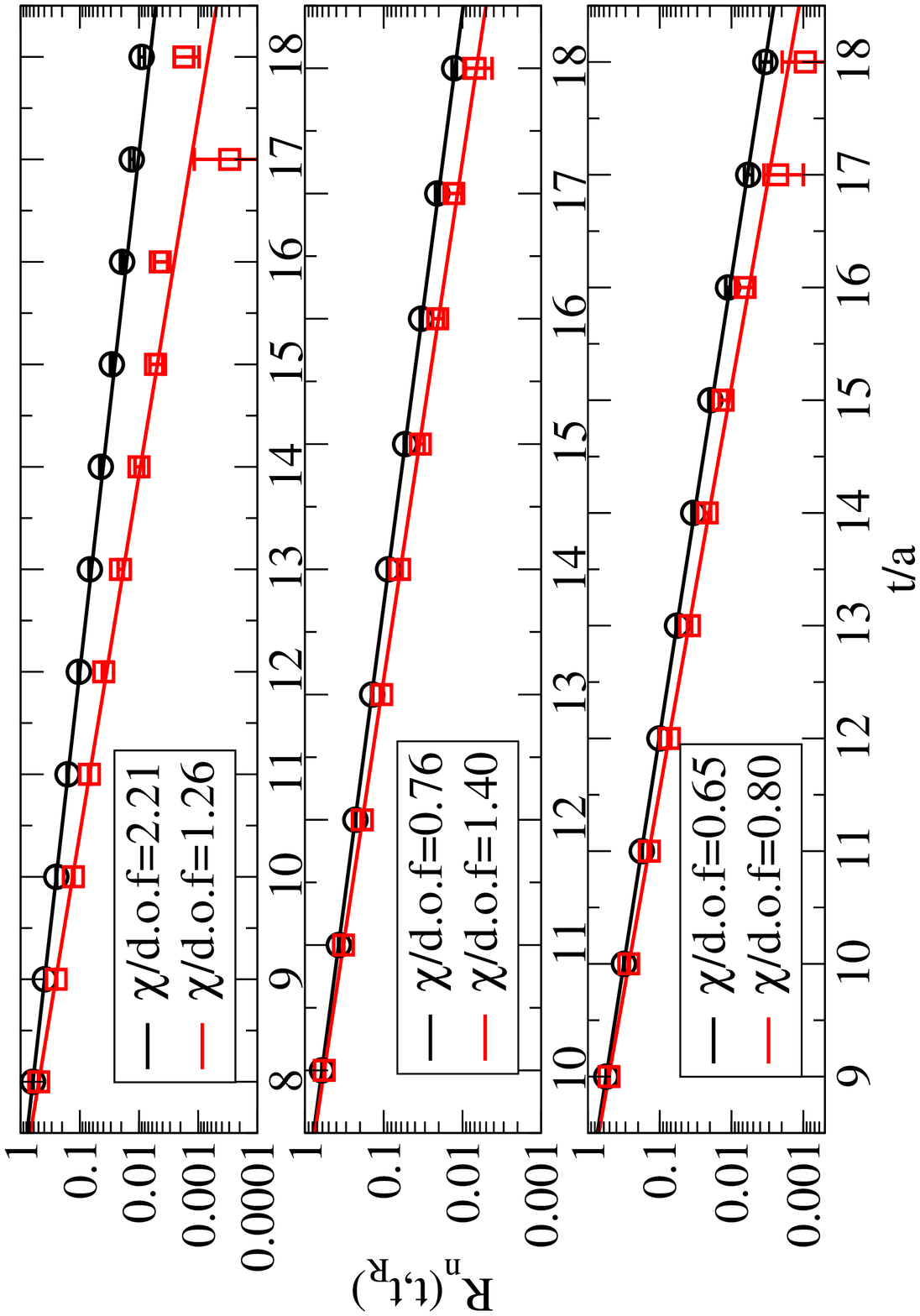}
\includegraphics[width=150pt,angle=\plotangle]{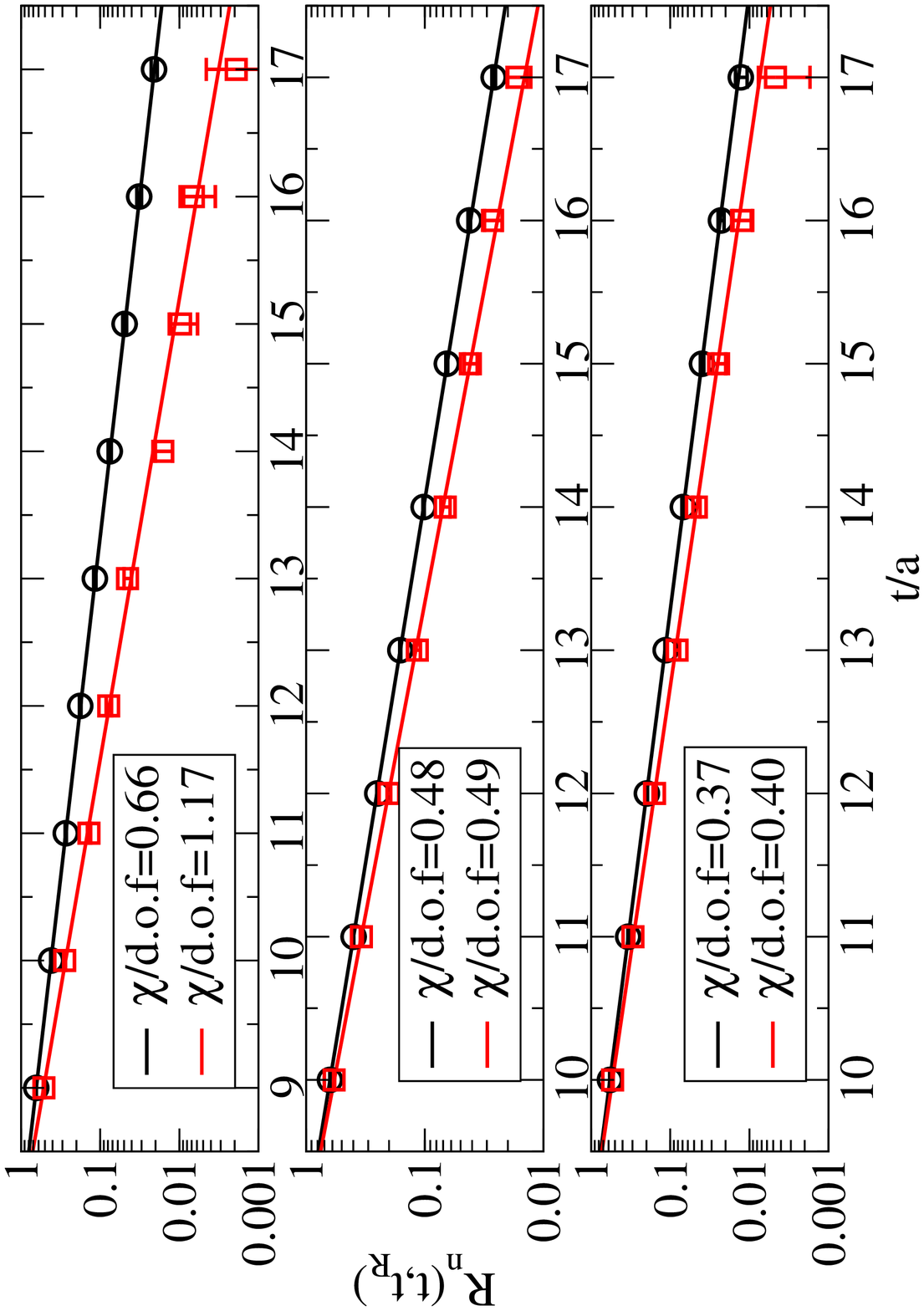}
\includegraphics[width=150pt,angle=\plotangle]{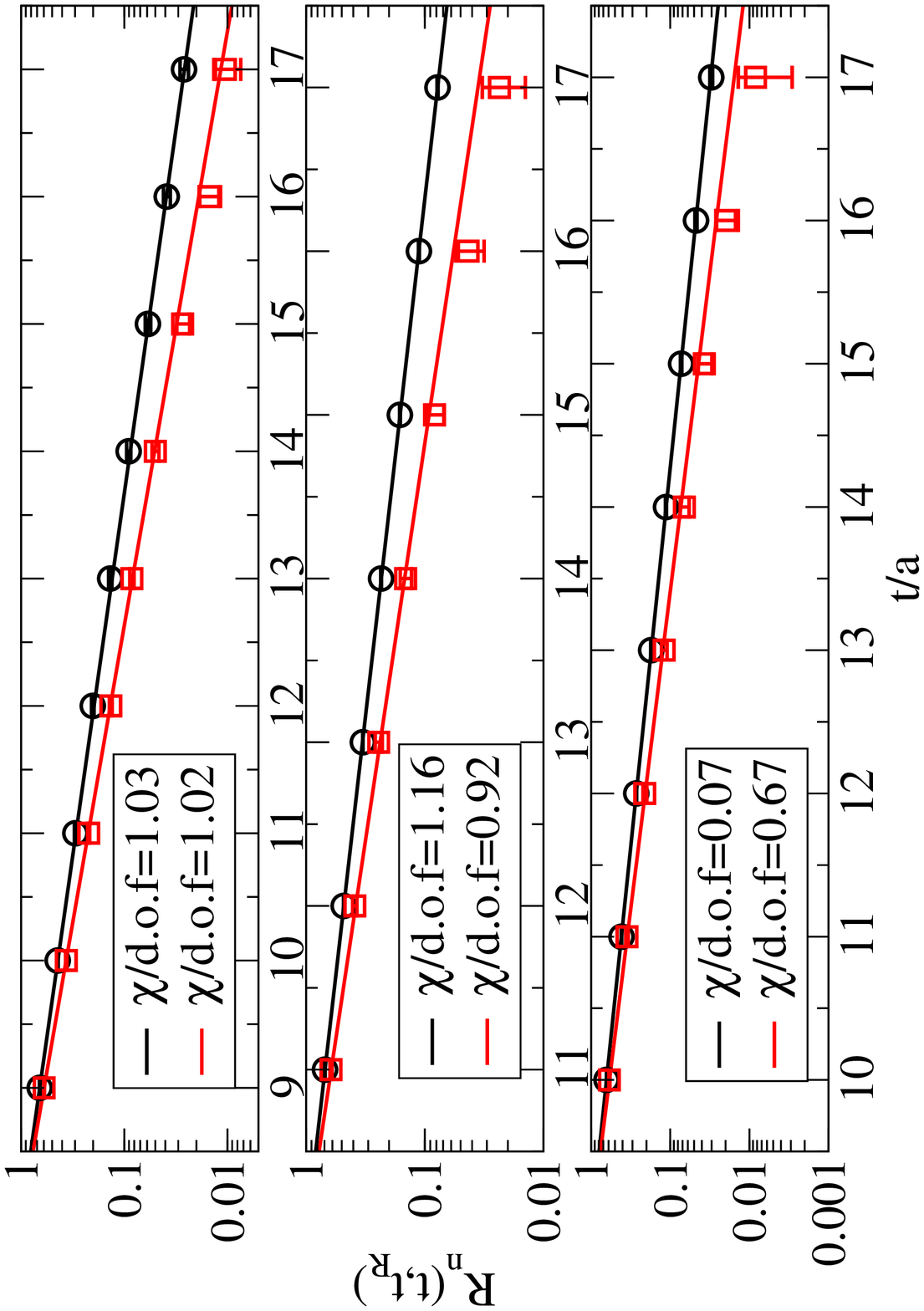}
\includegraphics[width=150pt,angle=\plotangle]{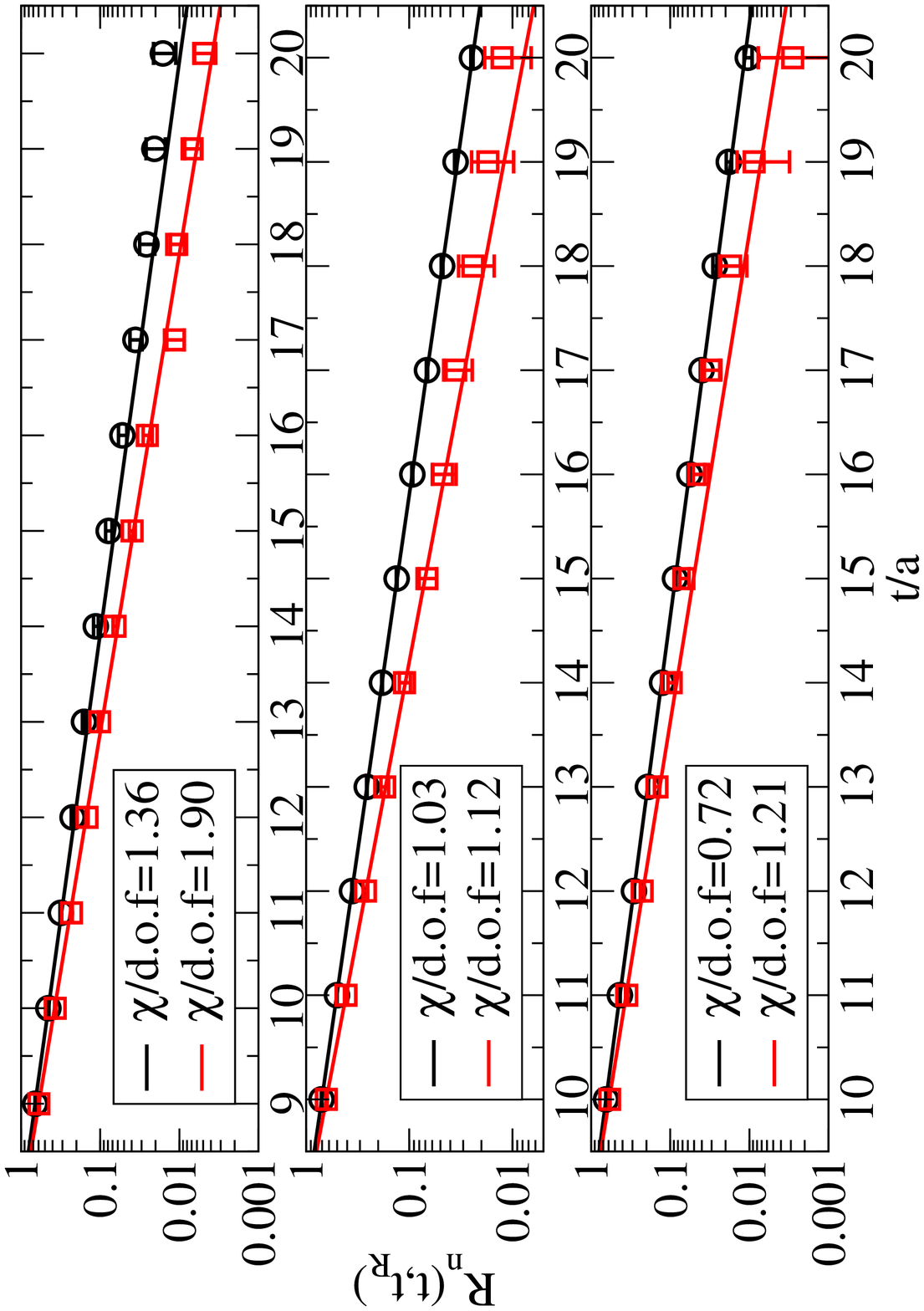}
%\end{tabular}
\end{center}
\caption{For the ensembles $A_1$ (upper left), $A_2$ (upper right), $A_3$ (lower left) and $A_4$ (lower right),
the correlator $R_n(t,t_R)$ ($n=1,2$)
as a function of $t$ is shown. For each ensemble, from top to bottom the three plots present
the lattice calculations in the CMF, MF1 and MF2, respectively. The solid
lines are correlated fits to Eq.~(\ref{eq:correlator_R_n}), from
which the energy eigenvalues $E_n$ are extracted. In each plot, the upper curve
is $n=1$ and the lower curve with the slightly steeper slope is $n=2$.}
\label{fig:correlator_A1}
\end{figure}

In Fig.~\ref{fig:correlator_A1} we show our
lattice results for $R_n(t,t_R)$ ($n=1,2$) in a logarithmic scale
for the CMF, MF1 and MF2, as a
function of time $t$ together with a correlated fit
to the asymptotic form given in
Eq.~(\ref{eq:correlator_R_n}).
From these fits we then extract the energies that will be used to determine
the scattering phase. Note that the slopes of $\ln(R_n(t,t_R))$ are often
very similar for $n=1$ and $n=2$, indicating that it is indeed essential
to use the correlation function matrix.
In order to extract the energies, we have to consider
the two
main sources of systematic error. One originates from the higher
excited states and affects the correlator in the low-$t$
region. The other arises from the unwanted thermal contributions
that distort the correlator in the large-$t$ region. By defining
a fitting window $[t_{\mathrm{min}},t_{\mathrm{max}}]$ and varying the
values of
$t_{\mathrm{min}}$ and $t_{\mathrm{max}}$,
we are able to control these systematic effects.
In practice, we set $t_\mathrm{min}$ to
be $t_R+1$ and increase the reference time slice $t_R$ to reduce the
higher excited state contaminations. Besides this, we set
$t_\mathrm{max}$ to be sufficiently far away from the time slice $t=T/2$
in order that the fitting results are protected from the unwanted
thermal contributions. The corresponding parameters $t_R$,
$t_{\mathrm{min}}$ and $t_{\mathrm{max}}$ used in this work are
listed in Table~\ref{tab:fitting_results}. All the ensembles shown
in Fig.~\ref{fig:correlator_A1} visibly
agree with the corresponding fit and lead to reasonable values of
$\chi^2/\mathrm{dof}$. The $\chi^2/\mathrm{dof}$
together with the fit results for $E_n$ ($n=1,2$) are also given
in Table~{\ref{tab:fitting_results}}.

\subsection{Lattice discretization effects}
In the continuum limit, the $E_{CM}$ is simply
related to the energy spectrum $E_n$  through the Lorentz
transformation of Eq.~(\ref{eq:Lorentz_cont}). However, on the lattice, the
discretization effects explicitly break Lorentz symmetry and
Eq.~(\ref{eq:Lorentz_cont}) is only valid up to
discretization errors. Another
discretization error arises from the continuum dispersion relation
in Eq.~(\ref{eq:continuum_dis_rel}),
which is particularly relevant for the finite-size methods used here.

 These two sources
 of systematic error have been studied in
 Ref.~\cite{Rummukainen:1995vs}, where the authors suggest to use the lattice modified relations
 \ba
 \label{eq:dispersion_lattice}
 &&\cosh(E_{CM})=\cosh(E_n)-2\sum_i\sin^2(P_i/2)\;,\quad n=1,2\;,\nn\\
 &&\cosh(E_{CM}/2)=2\sin^2(p^*/2)+\cosh(m_\pi)\;,\quad p^*=(2\pi/L)q\;,
 \ea
 instead of the continuum relations to reduce these
discretization errors. Following this suggestion, we
 calculate the
 energy $E_{CM}$ and the momentum $p^*$ from the energy eigenvalues $E_n$ using
 Eq.~(\ref{eq:dispersion_lattice})
 and then estimate the P-wave scattering phase $\delta_1$ by
 employing $p^*$ in the finite-size formulae.
 The results for $E_{CM}$, $p^*$ and $\delta_1$ are given in
 Table~{\ref{tab:rho_decay_results}}.

\subsection{Extraction of resonance parameters}
\begin{figure}
\begin{center}
%\begin{tabular}{cc}
\includegraphics[width=150pt,angle=\plotangle]{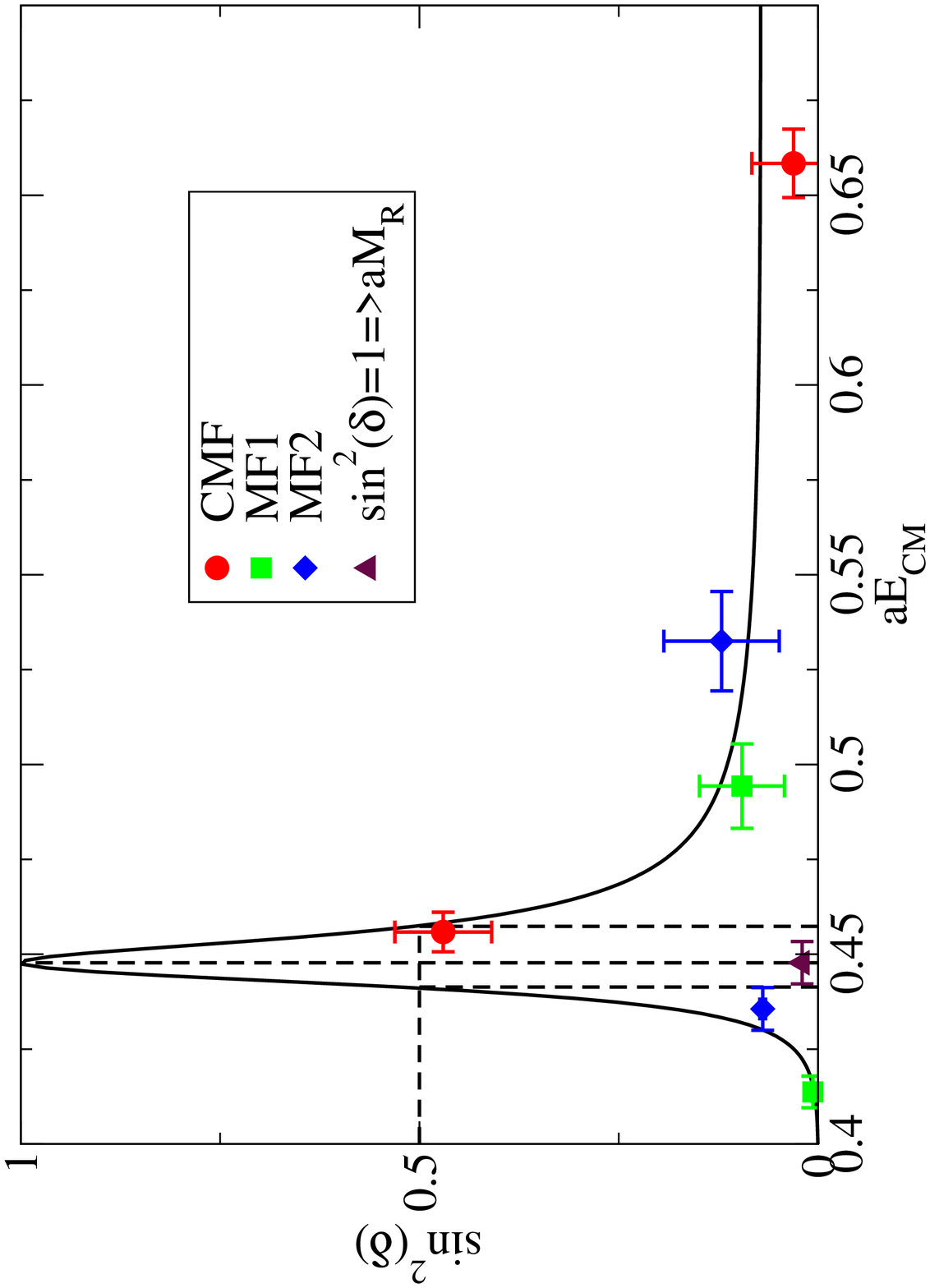}
\includegraphics[width=150pt,angle=\plotangle]{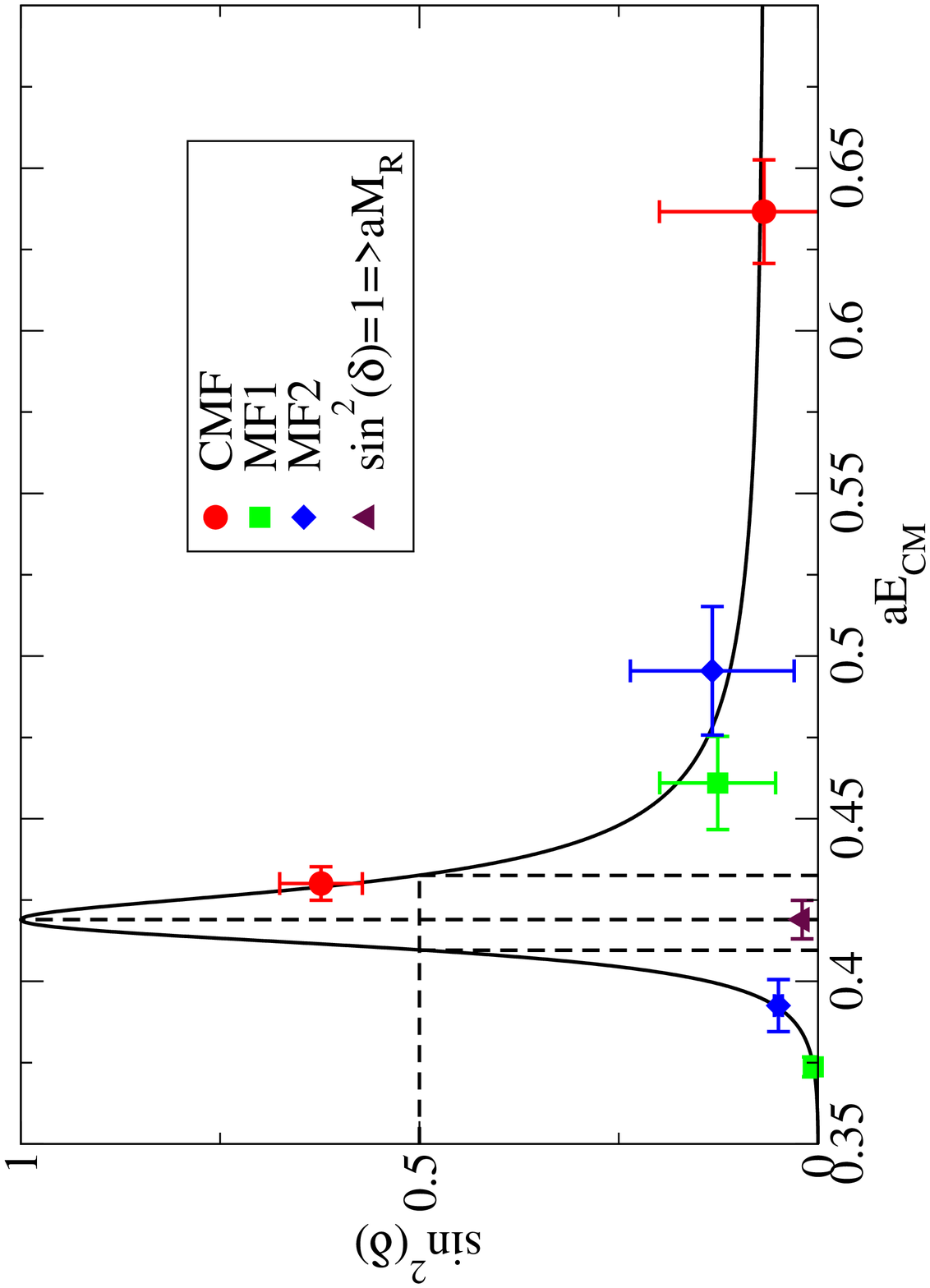}
\includegraphics[width=150pt,angle=\plotangle]{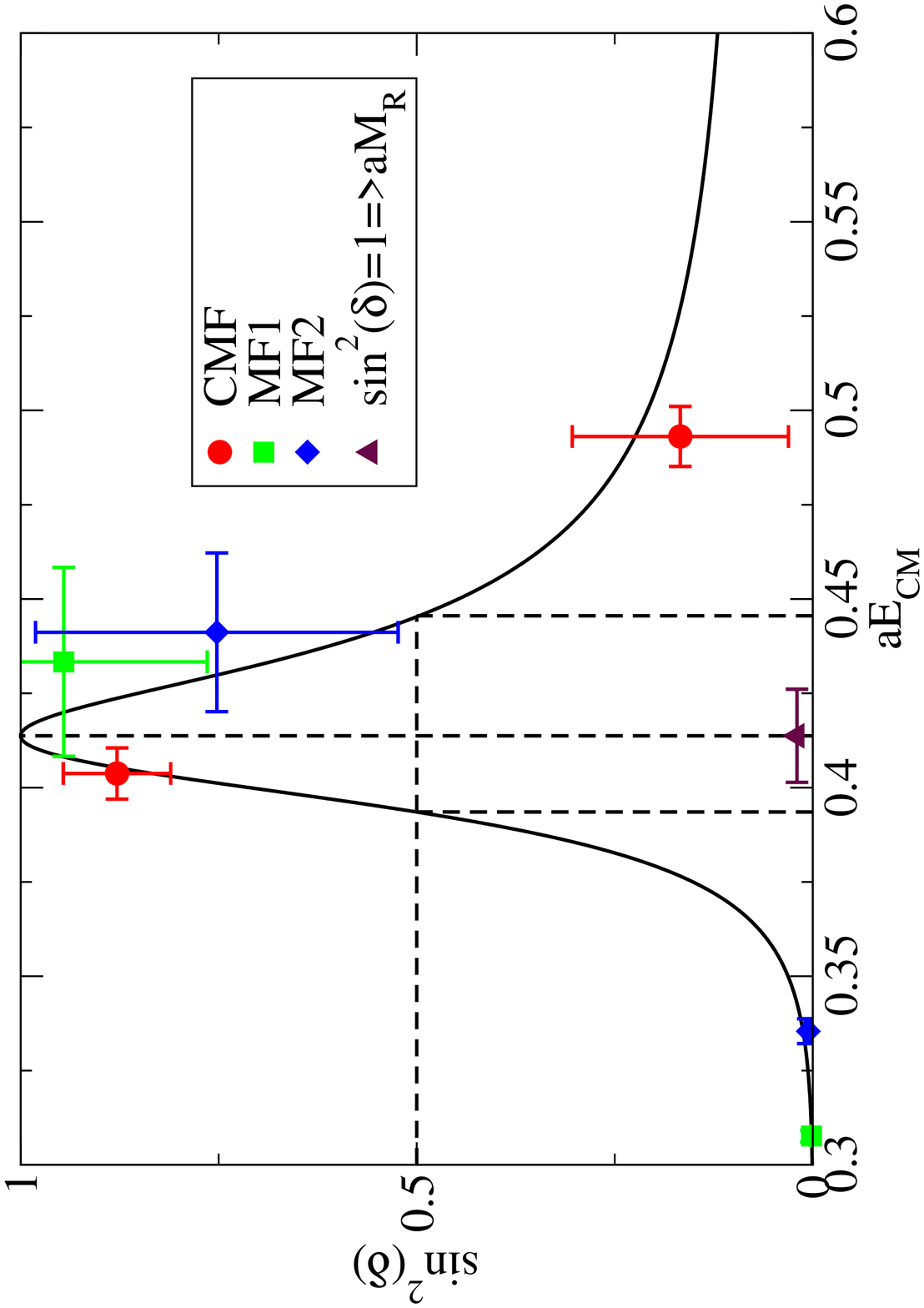}
\includegraphics[width=150pt,angle=\plotangle]{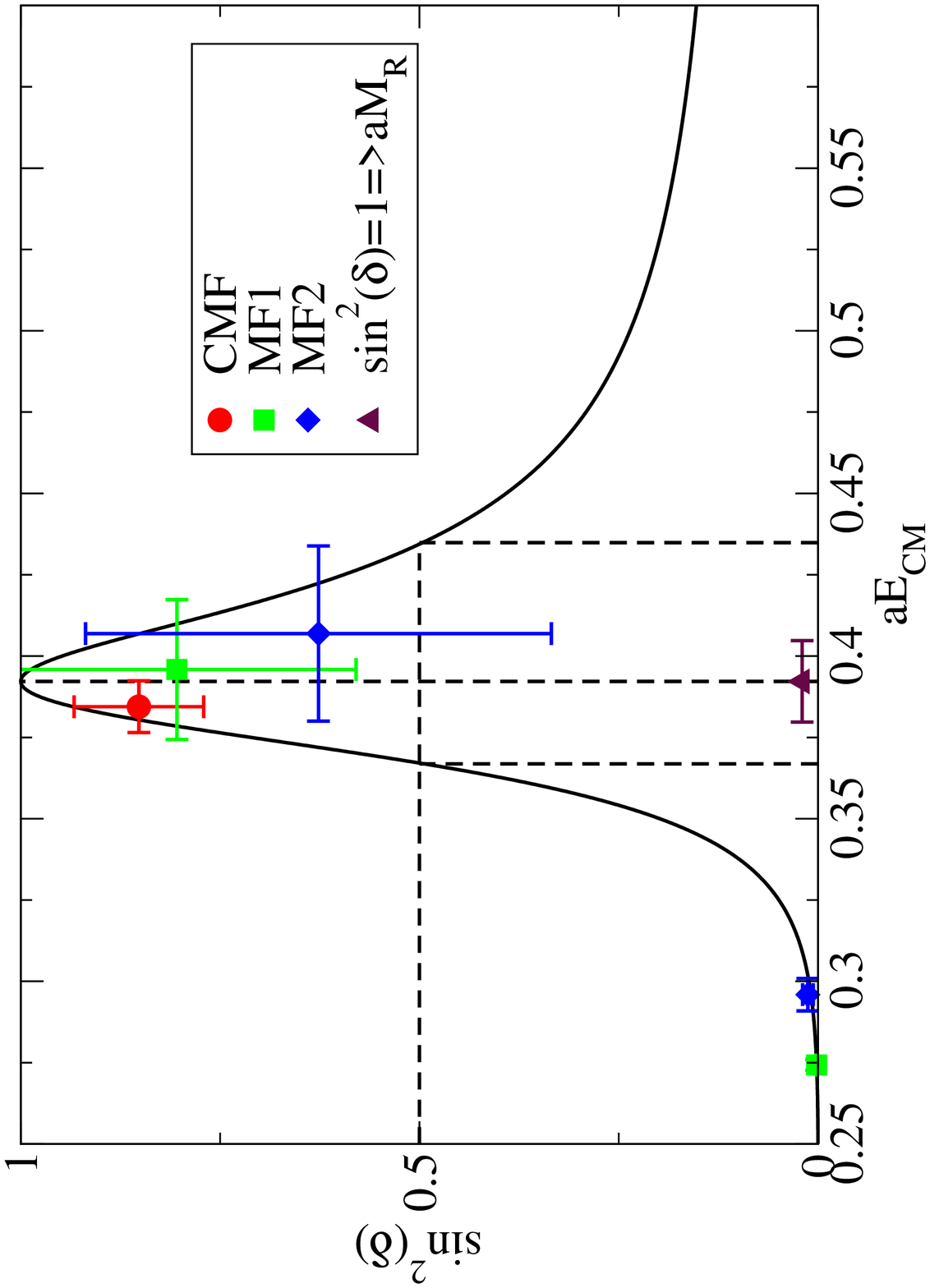}
%\end{tabular}
\end{center}
\caption{We show for the ensembles $A_1$ (upper left), $A_2$ (upper right), $A_3$ (lower left) and
 $A_4$ (lower right), the scattering phases calculated in the
CMF, MF1 and MF2 together with the fits to the effective range formula
Eq.~(\ref{eq:effective_range_formula}). At
the position where the scattering phase passes $\pi/2$,
the resonance mass $m_\rho$ (denoted as $aM_R$ in the graph) is
determined.  Through the fit, the coupling constant
$g_{\rho\pi\pi}$ and decay width $\Gamma_\rho$ are also extracted.}
\label{fig:scattering_phase_A1}
\end{figure}

From the $E_{CM}$ we can now compute
the P-wave scattering phases from six different
energy levels, two from each of the three Lorentz frames employed.
In order to extract the $\rho$-meson resonance parameters,
we fit the results for the scattering phase to the effective range
formula Eq.~(\ref{eq:effective_range_formula}) and show the corresponding fits in
Fig.~\ref{fig:scattering_phase_A1}.
At the position where the scattering phase passes $\pi/2$, the
resonance mass $m_\rho$ is determined. Additionally, the values of
$g_{\rho\pi\pi}$ and hence $\Gamma_\rho$ are also evaluated from the
fit. The corresponding results are given in
Table~\ref{tab:rho_mass_width}.
\begin{table}[tb]
\begin{center}
\begin{tabular}{|c|c|c|c|c|}
\hline  & $m_\pi$ (MeV) &$m_\rho$ (MeV) & $\Gamma_\rho$ (MeV) & $g_{\rho\pi\pi}$ \\
% \hline 3.90  & 0.0100 & $24$ & 520 & 2.77(2) & 479 & 7.23(59)(41) & -0.297(20)(16) \\
 \hline $A_1$ & 480 & 1118(14) & 39.5(8.2) & 6.46(40) \\
 \hline $A_2$ & 420 & 1047(15) & 55(11) & 6.19(42)\\
 \hline $A_3$ & 330 & 1033(31) & 123(43) & 6.31(87)\\
 \hline $A_4$ & 290 & 980(31) & 156(41) & 6.77(67)\\
% \hline $B_1$ & 320 & 997(52) & 158(68) & 7.3(1.1)\\
 \hline
\end{tabular}
\end{center}
% This is a hack that we should fix somehow.
\vspace{10pt} \caption{The results for the $\rho$-meson mass $m_\rho$,
the decay width $\Gamma_\rho$ and the effective $\rho\rightarrow\pi\pi$ coupling
$g_{\rho\pi\pi}$ at pion masses ranging from 480 to 290~MeV.}
\label{tab:rho_mass_width}
\end{table}

%Here, we want to add a word of caution.
The finite-size methods are valid only for
elastic scattering processes. In a situation with
large enough energy, i.e.\ when $E_{CM}>4m_\pi$, the inelastic
scattering channel will open and it is unclear how to
the determine the scattering phase in such a case. Therefore, in our
calculations we exclude results with energy $E_{CM}\gtrsim
4m_\pi$, which happened, fortunately, only for
the excited state in the CMF for ensemble $A_4$.

\subsection{Comparison with other results} 
\begin{figure}[hbt]
\begin{center}
\hspace{-15pt}\includegraphics[width=280pt,angle=\plotangle]{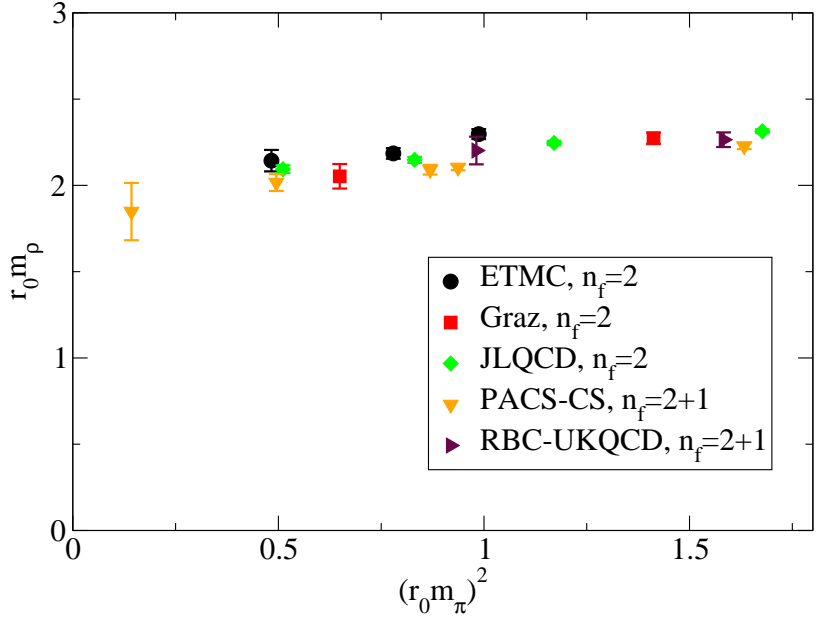}
\end{center}
\caption{$\rho$-meson mass as function of the pion mass squared, both scaled with $r_0$. 
The $\rho$-meson resonance masses determined in our calculations (ETMC) are compared with those of 
the groups listed in the legend:\ chirally improved fermions (Graz)~\cite{Gattringer:2008vj}, 
overlap fermions (JLQCD)~\cite{Aoki:2009qn,Aoki:2008tq}, 
%domain wall fermions on rooted staggered sea quarks (LHP)~\cite{WalkerLoud:2008bp}, 
%rooted staggered fermions (MILC)~\cite{Aubin:2004wf,Aubin:2006xv}, 
nonperturbatively improved Wilson fermions (PACS-CS)~\cite{Aoki:2008sm} and
domain wall fermions (RBC-UKQCD)~\cite{Allton:2007hx,Li:2006gra}.
In order to be consistent, we include only the results of those groups
for which we could readily find the values of $r_0$ evaluated at the same coupling and pion
mass as is the $\rho$-meson mass.  Also, note that only our calculation
includes a proper treatment of the resonance nature of the $\rho$-meson.}
\label{fig:rho_mass_compare}
\end{figure}

Using the resonance masses determined in the previous section, we show our values for $m_\rho$
together with 
those of other groups in Fig.~\ref{fig:rho_mass_compare} as a function of $m_\pi$. 
In order to compare these results, 
we scale $m_\rho$ and $m_\pi$ with the Sommer scale $r_0$~\cite{Sommer:1993ce} as determined 
by the groups individually. 
This avoids systematic effects when determining the lattice spacing from 
different observables and is most appropriate when one aims only at a comparison
of results between different groups. 
We find a rather satisfactory agreement and  
attribute the mild variation among the groups
with possible residual cutoff and finite-size effects in the various calculations, 
although a definite conclusion cannot be given here.  

We remark that 
our values of $m_\rho$ in physical units result from using the lattice spacing a=0.079~fm 
given earlier. 
This value of the lattice spacing was determined in Ref.~\cite{Baron:2009wt} by 
fixing the physical value 
of the pion decay constant $f_\pi$. 
%Other choices are made by other groups, resulting 
%in different lattice artifacts in different observables. This is an overall scale setting 
%that is not relevant to the methods described here to extract resonance parameters.

\subsection{Quark mass dependence}
Having analyzed the ensembles listed in Table~\ref{tab:ensemble1}
allows us to discuss now
the quark mass dependence of the $\rho$-meson resonance
parameters. There are several works using effective field theory (EFT)
to describe the quark mass dependence of the
$\rho$-meson resonance parameters
\cite{Jenkins:1995vb,Bijnens:1997ni,Leinweber:2001ac,Bruns:2004tj,Hanhart:2008mx}.
The general structure of the pion mass dependence of
$m_\rho$ and $\Gamma_\rho$ can be written down as
 \ba
  m_\rho&=&M_\rho^0+C_{m1}
  M_\pi^2+C_{m2}M_\pi^3+O(M_\pi^4)\;,\nn\\
 \Gamma_\rho&=&\Gamma_\rho^0+C_{\Gamma1}M_\pi^2+C_{\Gamma2}M_\pi^3+O(M_\pi^4)\;.\nn
 \ea
However, before using these formulae, it should be realized
that $m_\rho$ and $\Gamma_\rho$ are not only statistically correlated, but
also inherently
related to each other, suggesting that the
coefficients $C_{mi}$ and $C_{\Gamma i}$ (i=1,2) are not independent
from each other.
Therefore, in this work we will follow the strategy of Refs.~\cite{Djukanovic:2009zn,Djukanovic:2010id}
where $m_\rho$ and
$\Gamma_\rho$ are considered as the real and
imaginary part of the complex pole of the $\rho$-meson propagator.
Hence, we
introduce the complex pole parameter $Z$ defined
through
 \bd
 Z=(m_\rho-i\Gamma_\rho/2)^2\;.
 \ed
 In this approach the power counting is given by
 the complex-mass renormalization scheme. 
Up to $O(q^4)$ in the chiral expansion, where $q$ is a typical pion momentum, 
$Z$ is given by~\cite{Djukanovic:2009zn,Djukanovic:2010id}
 \ba
 Z&=&Z_{\chi}+C_{\chi}M_\pi^2-\frac{g^2_{\omega\rho\pi}}{24\pi}Z_{\chi}^{1/2}M_\pi^3\nn\\
 \label{eq:z_expansion}
 &&-\frac{g^2_{\omega\rho\pi}}{32\pi^2}M_\pi^4\left(\ln\frac{M_\pi^2}{M_{\chi}^2}-1\right)
 +\frac{g^2}{16\pi^2}\frac{M_\pi^4}{M_{\chi}^2}\left(3-2\ln\frac{M_\pi^2}{M_{\chi}^2}-2i\pi\right)\;,
 \ea
 where
 $Z_{\chi}=(M_{\chi}-i\Gamma_{\chi}/2)^2$ is the pole of the 
$\rho$-meson propagator in the chiral limit, 
 $M_\pi^2$ is the lowest-order expression of the chiral expansion for the squared pion mass
 and $C_{\chi}$, $g_{\omega\rho\pi}$ and $g$ are coupling constants assuming real values.
 Using Eq.~(\ref{eq:z_expansion}) to fit our results,
we can determine the value of $Z$ at the physical
 point, where it can be converted to the physical resonance mass
$m_{\rho,\mathrm{phy}}$ and decay width
$\Gamma_{\rho,\mathrm{phy}}$.

 In practice, we perform the chiral extrapolation of $Z$ in terms of the pion mass $m_\pi$
as extracted from the pseudoscalar correlation function as measured 
directly in the numerical 
calculations.
By inserting the relation
 \bd
 m_\pi^2=M_\pi^2\left\{1+\frac{M_\pi^2}{32\pi^2F_\pi^2}\ln\frac{M_\pi^2}{\Lambda_3^2}+O(M_\pi^4)\right\}
 \ed
 into Eq.~(\ref{eq:z_expansion}), the expression for $Z$ in terms of $m_\pi$ is given by
 \ba
 Z&=&Z_{\chi}+C_{\chi}m_\pi^2-\frac{C_{\chi}m_\pi^4}{32\pi^2F_\pi^2}\ln\frac{m_\pi^2}{\Lambda_3^2}
 -\frac{g^2_{\omega\rho\pi}}{24\pi}Z_{\chi}^{1/2}m_\pi^3\nn\\
 \label{eq:z_expansion_new}
 &&-\frac{g^2_{\omega\rho\pi}}{32\pi^2}m_\pi^4\left(\ln\frac{m_\pi^2}{M_{\chi}^2}-1\right)
 +\frac{g^2}{16\pi^2}\frac{m_\pi^4}{M_{\chi}^2}\left(3-2\ln\frac{m_\pi^2}{M_{\chi}^2}-2i\pi\right)\;.
 \ea
 In Eq.~(\ref{eq:z_expansion_new}) the values of the input parameters $F_\pi$ and $\Lambda_3$ are taken from 
 Ref.~\cite{Baron:2009wt} with
 \bd
 F_\pi=\frac{1}{\sqrt{2}}f_0=86(1)~\mev\;,\quad \ln\left(\Lambda_3^2/m^2_{\pi,\mathrm{phy}}\right)=\bar{l}_3=3.50(31)\;,
 \ed
 where $m_{\pi,\mathrm{phy}}$ is the physical pion mass.

% It is important to note that extrapolating in $m_\pi$ does indeed change
% the expression for $Z$ but only at the order of $M_\pi^4\ln{M_\pi^2}$.
% $M_\pi^2=2B_0\mu$ to the physical point of $m_{\pi,\mathrm{phy}}^2=2B_0\mu_{\pi}$, where $\mu$ is the bare quark mass
% listed in Table~\ref{tab:ensemble1}. The values of
% parameter $B_0$ and $\mu_\pi$ are taken from Ref.~\cite{Boucaud:2007uk}
% \bd
% 2B_0\mu=4.99(6)\;,\quad a\mu_{\pi}=0.00078(2)\;.
% \ed
% We also extrapolate $Z$ in terms of the lattice-physical pion mass $m_\pi$. By lattice-physical we mean the pion mass
% as measured from the correlation function. It is important to note that extrapolating in $m_\pi$ does indeed change
% the expression for $Z$. By assuming such change is small, we find very consistant results.

 Before we perform a precise test of Eq.~(\ref{eq:z_expansion_new}),
we first confront our lattice results with a simplified fit ansatz to order $O(q^3)$, namely
 \be
 \label{eq:fit_M3}
 Z=Z_{\chi}+C_{\chi}m_\pi^2-\frac{g^2_{\omega\rho\pi}}{24\pi}Z_{\chi}^{1/2}m_\pi^3\;.
 \ee
% where $C_3$ is a real parameter, whose value is suggested
% by EFT as $-g_{\omega\rho\pi}^2/24\pi=-3.4$~$\mathrm{GeV}^{-2}$~\cite{Leinweber:2001ac}. However since the value 
% of $C_3$ in Eq.~(\ref{eq:fit_M3}) is commonly treated as a fitting parameter,
% we are not guranteed that it has the same value required by EFT.

 \begin{table}[htb]
 \begin{center}
 \begin{tabular}{|c|c|c|}
 \hline
 Fit of $Z$ to & Eq.~(\ref{eq:fit_M3}) &  Eq.~(\ref{eq:z_expansion_new})  \\
 \hline
 \hline
 $m_{\rho,\mathrm{phy}}$      & 0.821(24) & 0.850(35) \\
 \hline
 $\Gamma_{\rho,\mathrm{phy}}$ & 0.171(31) & 0.166(49)\\
 \hline
 \hline
 $M_\chi$                     & 0.756(24) & 0.803(47) \\
 \hline
 $\Gamma_\chi$                & 0.190(35) & 0.179(58) \\
 \hline
 $C_\chi$                     & 6.42(45)  & 4.9(2.0)\\
 \hline
 $g_{\omega\rho\pi}^2/24\pi$  & 9.8(1.5)  & 10(12)  \\
 \hline
 $g^2/(16\pi^2M_{\chi}^2)$    &   ---     & 0.01(1.09) \\
 \hline
 \end{tabular}
 \end{center}
 \vspace{10pt} \caption{The physical $\rho$-meson mass and decay width as extracted
using Eq.~(\ref{eq:fit_M3}) and Eq.~(\ref{eq:z_expansion_new}).
 The values of $m_{\rho,\mathrm{phy}}$, $\Gamma_{\rho,\mathrm{phy}}$,
$M_\chi$ and $\Gamma_\chi$ are
 given in units of GeV and $g_{\omega\rho\pi}^2/24\pi$ and $g^2/(16\pi^2M_{\chi}^2)$ are in
 units of $\mathrm{GeV}^{-2}$.}
 \label{tab:diff_fit_model}
 \end{table}

\begin{figure}[hbt]
\begin{center}
\hspace{-15pt}\includegraphics[width=280pt,angle=\plotangle]{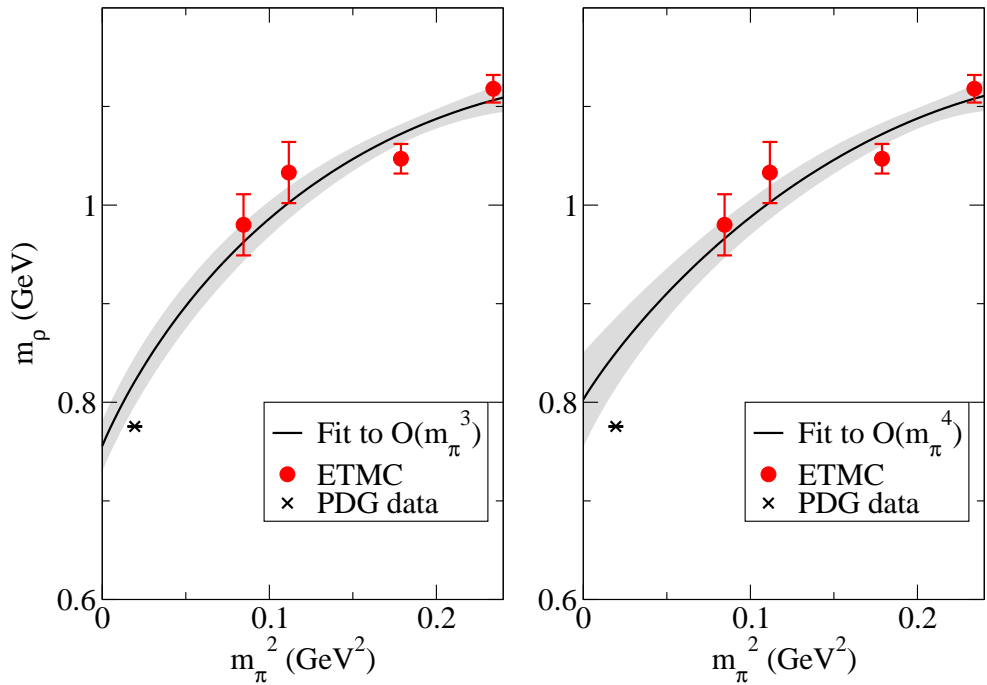}
\end{center}
\caption{The $\rho$-meson resonance mass as a function of
 the square of the pion mass. In the left panel, we fit the lattice results to Eq.~(\ref{eq:fit_M3}).
 In the right panel, we fit them to Eq.~(\ref{eq:z_expansion_new}).  Note that these are
combined fits to $m_\rho$ and $\Gamma_\rho$ (shown in Fig.~\ref{fig:width_R}).}
\label{fig:M_R}
\end{figure}

 In the left panel of Fig.~\ref{fig:M_R} we plot the mass of the $\rho$-meson as a function of
 the square of the pion mass together with a fit to Eq.~(\ref{eq:fit_M3}).
 Using the fit to extrapolate to the physical point, our lattice result
 turns out to lie slightly high relative to the PDG value of the $\rho$-meson mass 
and shows a deviation of $1.9\,\sigma$.

 In order to see whether higher-order corrections could reconcile
our calculation with the experimentally determined $\rho$-meson mass,
we also fit our lattice results to Eq.~(\ref{eq:z_expansion_new}). 
All the fit results are listed in
 Table~\ref{tab:diff_fit_model}. From the simplified fit to Eq.~(\ref{eq:fit_M3}) 
the lattice result of $g_{\omega\rho\pi}^2/24\pi$ 
 is larger than the one suggested 
 by EFT, which is $g_{\omega\rho\pi}^2/24\pi=3.4$~$\mathrm{GeV}^{-2}$~\cite{Leinweber:2001ac}. 
After including the
 terms of $O(q^4)$ in the fit, the uncertainty of the determination of 
$g_{\omega\rho\pi}^2/24\pi$ 
 becomes, unfortunately, much larger and in fact $g_{\omega\rho\pi}^2/24\pi$ 
cannot be determined in a statistically significant way.
% As a result the value of $g_{\omega\rho\pi}^2/24\pi$ as extracted using a fit to 
%Eq.~(\ref{eq:z_expansion_new}) 
%statistically agrees with the one given by EFT. 
A similar
 situation happens in the determination of the parameter $g^2/(16\pi^2M_{\chi}^2)$. 
 The KSFR relation~\cite{Kawarabayashi:1966kd,Riazuddin:1966sw}
 \bd
 M_{\chi}^2=2g^2F_\pi^2
 \ed
 suggests that $g^2/(16\pi^2M_{\chi}^2)$ takes the value of 
$1/(32\pi^2F_\pi^2)=0.43$~$\mathrm{GeV}^{-2}$.  However, we are unable to determine $g^2/(16\pi^2M_{\chi}^2)$ reliably from the fit.
As can be inferred from Table~\ref{tab:diff_fit_model}, using a fit to 
Eq.~(\ref{eq:z_expansion_new}) there is a 
 40\% uncertainty in the determination of $C_{\chi}$ and a more than 
 100\% uncertainty in the determinations of both 
$g_{\omega\rho\pi}^2/24\pi$ and $g^2/(16\pi^2M_{\chi}^2)$.  Proceeding with these results, 
 we plot the mass of the $\rho$-meson as a function of
 the square of the pion mass together with the
 fit to Eq.~(\ref{eq:z_expansion_new}) in the right panel of Fig.~\ref{fig:M_R}.
 At the physical point the result of $m_\rho$ is still high relative to the PDG
 value, suggesting that the pion masses used in the current calculations
 are too high for even the $O(q^4)$ extrapolations and that 
yet lighter quark masses will be necessary for quantitatively precise comparisons
with experimental measurements of the $\rho$-meson mass.

%Note that these extrapolations are dominated by the data of resonance parameters 
% obtained at pion mass $m_\pi\gtrsim300$~MeV. To fully incorporate the chiral dynamics of these extrapolations
% the study of the $\rho$--resonance at yet lighter quark masses is very necessary. Besides this, the statistical errors
% should also be well reduced to include the $O(m_\pi^4)$ contributions reliably in the fit. 

\begin{figure}[ht]
\begin{center}
\hspace{-15pt}\includegraphics[width=280pt,angle=\plotangle]{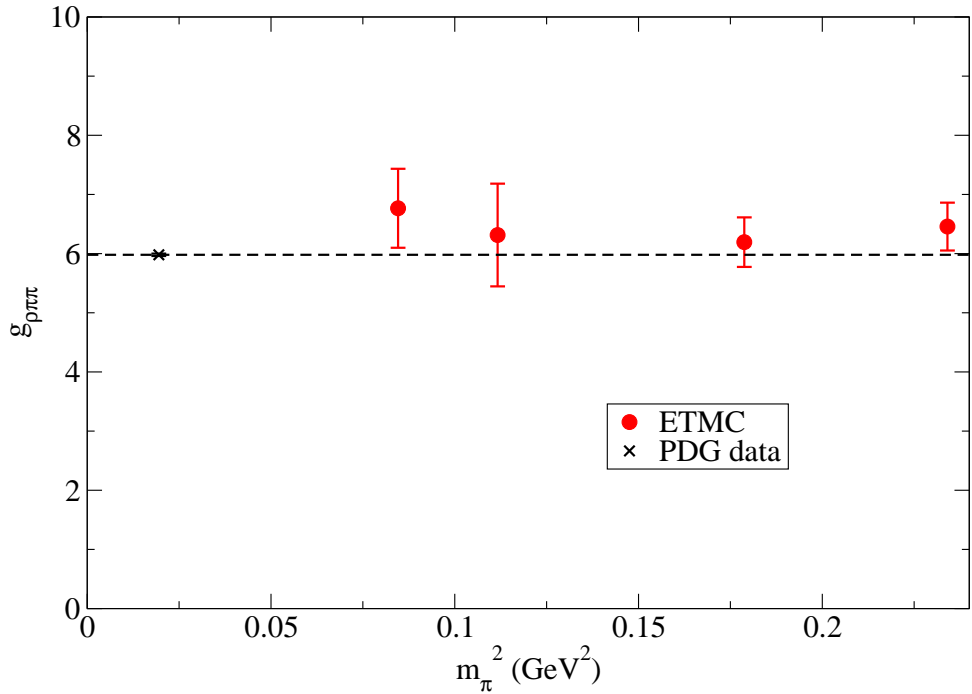}
\end{center}
\caption{The effective coupling $g_{\rho\pi\pi}$ as a function of
 the square of the pion mass.}
\label{fig:effective_coupling}
\end{figure}

 In Fig.~\ref{fig:effective_coupling}, we plot the coupling
 $g_{\rho\pi\pi}$ as a function of the square of the pion mass and
 find that $g_{\rho\pi\pi}$ is practically independent of the pion
 mass.  Moreover, the value of $g_{\rho\pi\pi}$ is consistent with the
 PDG value.  This is not entirely unexpected.  The coupling
 $g_{\rho\pi\pi}$, being dimensionless, is expected to be less
 sensitive to the pion masses and lattice spacings used in the
 calculation than the resonance mass $m_\rho$ is.  In fact, whereas
 the accuracy of $m_\rho$ is currently systematically limited by the
 pion masses used in the calculation, the precision with which we can
 calculate $g_{\rho\pi\pi}$ is clearly dominated by just the
 statistical errors of the current calculation.

\begin{figure}[ht]
\begin{center}
\hspace{-15pt}\includegraphics[width=280pt,angle=\plotangle]{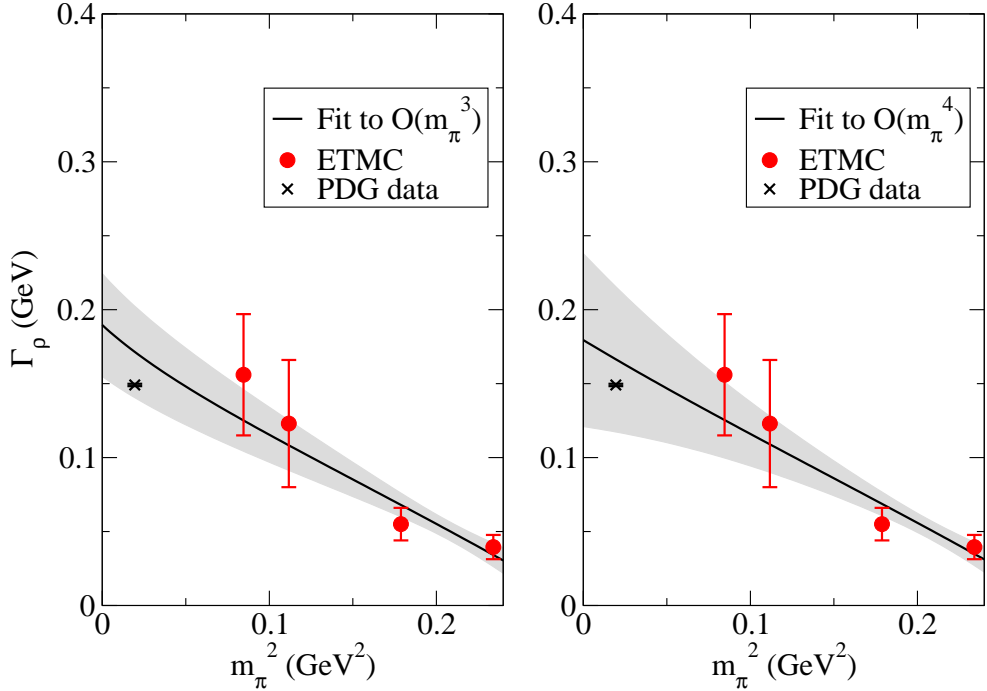}
\end{center}
\caption{The $\rho$-meson decay width as a function of
 the square of the pion mass. The left panel shows the lattice results and the fit to Eq.~(\ref{eq:fit_M3}).
 The right panel shows the fit to Eq.~(\ref{eq:z_expansion_new}).  Note that these are
combined fits to $\Gamma_\rho$ and $m_\rho$ (shown in Fig.~\ref{fig:M_R}).}
\label{fig:width_R}
\end{figure}

 Eq.~(\ref{eq:decay_width}) shows that the decay width is determined
 from the fitted values of both $m_\rho$ and $g_{\rho\pi\pi}$.  Hence,
 we expect that it will reflect a combination of the aspects just
 discussed.  In fact, in the chiral limit Eq.~(\ref{eq:decay_width})
 reduces to $\Gamma_\rho = m_\rho g^2_{\rho\pi\pi} / (48\pi)$.  Thus
 the fact that $m_\rho$ overshoots the experimental measurement
 implies that $\Gamma_\rho$ will also be larger than the measured
 value.  Additionally, the error of $g_{\rho\pi\pi}$ will be enhanced
 in $\Gamma_\rho$ leading to larger errors in the width than in the
 mass.  These features can indeed be seen in Fig.~\ref{fig:width_R},
 where we show the lattice results for $\Gamma_\rho$ as a function of
 the square of the pion mass together with the fit to
 Eq.~(\ref{eq:fit_M3}) in the left panel and with the fit to
 Eq.~(\ref{eq:z_expansion_new}) in the right panel.  At the physical
 point, the decay widths are obtained as
 $\Gamma_{\rho,\mathrm{phy}}=171(31)~\mev$ using the fit to
 Eq.~(\ref{eq:fit_M3}) and as
 $\Gamma_{\rho,\mathrm{phy}}=166(49)~\mev$ using the fit to
 Eq.~(\ref{eq:z_expansion_new}).  Both of the results are consistent
 with the PDG value $\Gamma_{\rho,\mathrm{PDG}}=149.1(0.8)~\mev$
 within 1$\sigma$.  Note, however, that obviously the values
 determined from our lattice calculation are much less accurate than
 the one extracted from experimental measurements.  Therefore, we
 consider the present work more as an initial study of how accurately
 resonance parameters can be extracted from nonperturbative lattice
 calculations and not as a precise determination of these
 parameters. The results we have obtained here demonstrate
 that resonances can indeed be analyzed on finite lattices with
 numerical calculations. This is very promising, given the number of
 hadrons that appear in the physical QCD spectrum as resonances.


\section{Conclusion}
In this work, we have calculated the P-wave
pion-pion scattering phase in the $I=1$ channel
near the $\rho$-meson resonance region. We have performed our calculations
at pion masses ranging from $480$ to $290$~MeV and at a lattice
spacing of $a=0.079$~fm. At all the pion masses, the physical
kinematics for the $\rho$-meson decay, $m_\pi/m_\rho<0.5$, is
satisfied. Compared to previous
calculations, we have pushed the techniques much farther forward
by employing three Lorentz frames simultaneously. This
allowed us, in particular, to map out the energy region of the resonance
without having to employ larger and more computationally demanding
lattice calculations.

Making use of L\"uscher's finite-size methods, we
evaluated the scattering phase from six energy eigenvalues
per ensemble. In this way, we
could fit the scattering phase with an effective range formula
allowing us to extract
the $\rho$-resonance mass $m_\rho$, the decay
width $\Gamma_\rho$ and the effective coupling $g_{\rho\pi\pi}$.
Taking the inherent relation between $m_\rho$ and $\Gamma_\rho$ into
account, we have performed a fit to our results, obtained
at four values of the pion mass, as a function of the complex parameter
$Z=(m_\rho-i\Gamma_\rho/2)^2$.
This provided a means of
extrapolation to the physical point.
Even though our fit formulae are guided by EFT,
our results are not precise enough to perform a thorough test
of the fit ans\"atze.
%including the higher order terms significantly reduces the accuracy
%of the extrapolation, especially in the low pion mass region. We found that EFT
%might not be fully adequate to describe the quark mass dependence of
%the $\rho$-meson mass and its decay width, even with pion masses
%as small as $m_\pi\sim300$~MeV.

Keeping in mind the caveats just discussed, we quote for the
$\rho$-meson mass $m_{\rho,\mathrm{phys}}=0.850(35)$~GeV and for the decay width
$\Gamma_{\rho,\mathrm{phys}}=0.166(49)$~GeV. When these values are compared to the
corresponding experimentally measured quantities, it is clear that
the lattice computations cannot yet match the experimental accuracy.
Although a precise determination of resonance parameters on the lattice is still a
challenge, our work serves as a next step in the attempt to understand the strong decays in a conceptually clean way.

\begin{acknowledgments}
This work is supported by the DFG Sonderforschungsbereich / Transregio SFB/TR-9 and 
the DFG Project No. Mu 757/13. X.\ F.\ is supported 
in part by the Grant-in-Aid of the Japanese Ministry of Education (No. 21674002) and
D.\ R.\ is supported in part by Jefferson Science Associates, LLC under U.S. DOE Contract No. DE-AC05-06OR23177.
 We thank G.~Herdoiza,
C.~Urbach and M.~Wagner for helpful suggestions and
assistance. X.\ F.\ would like to thank N. Ishizuka, C. Liu and C. Michael for valuable
discussions. The computer time for this project
was made available to us by the John von Neumann Institute for
Computing on the JUMP and JUGENE systems in J\"ulich. Part of the
analysis runs were performed in the computer centers of DESY Zeuthen
and the IDRIS (Paris-sud). We thank these computer centers and their staff for technical
support.
\end{acknowledgments}

\appendix
%%%%%%%%%%%%%%%%%%%%%%%%%%%%%%%%%%%%%%%%%%%%%%%%%%%%%%%%
\begin{table}[htb]
\begin{center}
\begin{tabular}{cccccccc}
 \hline                  & Frame & $t_R/a$ & $t_{\mathrm{min}}/a$ & $t_{\mathrm{max}}/a$ & n & $\chi^2/\mathrm{dof}$ & $aE_n$ \\
 \hline
 \hline                  &       &       &                    &                    & 1 & 2.21 & 0.4559(52) \\
                         & \raisebox{1.5ex}{CMF} & \raisebox{1.5ex}{7} &
                         \raisebox{1.5ex}{8} & \raisebox{1.5ex}{18} & 2 & 1.26 & 0.6584(90)                \\
                         &       &       &                    &                    & 1 & 0.76 & 0.4869(35) \\
  \raisebox{1.5ex}{$A_1$}& \raisebox{1.5ex}{MF1} & \raisebox{1.5ex}{9} &
                         \raisebox{1.5ex}{10} & \raisebox{1.5ex}{18} & 2 & 1.40 & 0.5563(98)               \\
                         &       &       &                    &                    & 1 & 0.65 & 0.5660(42) \\
                         & \raisebox{1.5ex}{MF2} & \raisebox{1.5ex}{8} &
                         \raisebox{1.5ex}{9} & \raisebox{1.5ex}{18} & 2 & 0.80 & 0.642(11)                 \\
 \hline
 \hline                  &       &       &                    &                    & 1 & 0.66 & 0.4301(52) \\
                         & \raisebox{1.5ex}{CMF} & \raisebox{1.5ex}{8} &
                         \raisebox{1.5ex}{9} & \raisebox{1.5ex}{17} & 2 & 1.17 & 0.637(16)                 \\
                         &       &       &                    &                    & 1 & 0.48 & 0.4537(25) \\
  \raisebox{1.5ex}{$A_2$}& \raisebox{1.5ex}{MF1} & \raisebox{1.5ex}{9} &
                         \raisebox{1.5ex}{10} & \raisebox{1.5ex}{17} & 2 & 0.49 & 0.527(12)                \\
                         &       &       &                    &                    & 1 & 0.37 & 0.5343(57) \\
                         & \raisebox{1.5ex}{MF2} & \raisebox{1.5ex}{9} &
                         \raisebox{1.5ex}{10} & \raisebox{1.5ex}{17} & 2 & 0.40 & 0.612(16)                \\
 \hline
 \hline                  &       &       &                    &                    & 1 & 1.03 & 0.4037(68) \\
                         & \raisebox{1.5ex}{CMF} & \raisebox{1.5ex}{8} &
                         \raisebox{1.5ex}{9} & \raisebox{1.5ex}{17} & 2 & 1.02 & 0.4931(80)                \\
                         &       &       &                    &                    & 1 & 1.16 & 0.3638(13) \\
  \raisebox{1.5ex}{$A_3$}& \raisebox{1.5ex}{MF1} & \raisebox{1.5ex}{10} &
                         \raisebox{1.5ex}{11} & \raisebox{1.5ex}{17} & 2 & 0.92 & 0.474(23)                \\
                         &       &       &                    &                    & 1 & 0.07 & 0.4330(25) \\
                         & \raisebox{1.5ex}{MF2} & \raisebox{1.5ex}{9} &
                         \raisebox{1.5ex}{10} & \raisebox{1.5ex}{17} & 2 & 0.67 & 0.518(18)                \\
 \hline
 \hline                  &       &       &                    &                    & 1 & 1.36 & 0.3844(79) \\
                         & \raisebox{1.5ex}{CMF} & \raisebox{1.5ex}{8} &
                         \raisebox{1.5ex}{9} & \raisebox{1.5ex}{20} & 2 & 1.90 & 0.4591(86)                \\
                         &       &       &                    &                    & 1 & 1.03 & 0.3363(14) \\
  \raisebox{1.5ex}{$A_4$}& \raisebox{1.5ex}{MF1} & \raisebox{1.5ex}{9} &
                         \raisebox{1.5ex}{10} & \raisebox{1.5ex}{20} & 2 & 1.12 & 0.440(19)                \\
                         &       &       &                    &                    & 1 & 0.72 & 0.4035(36) \\
                         & \raisebox{1.5ex}{MF2} & \raisebox{1.5ex}{9} &
                         \raisebox{1.5ex}{10} & \raisebox{1.5ex}{20} & 2 & 1.21 & 0.490(22)                \\
 \hline
\end{tabular}
\end{center}
\caption{Values of the energy eigenvalues for the ground
state ($n=1$) and the first excited state ($n=2$)
in the CMF, MF1 and MF2. In the
table we list the ensemble number, the reference time $t_R$, the
lower and upper bound of the fitting window, $t_{\mathrm{min}}$ and
$t_{\mathrm{max}}$, the fit quality $\chi^2/\mathrm{dof}$ and
the fit results for energy eigenvalues $E_n$ ($n=1,2$).}
\label{tab:fitting_results}
\end{table}

\begin{table}[htb]
\begin{center}
\begin{tabular}{cccccccccc}
 \hline & Frame & $n$ & $aE_n$ & $aE_{CM}$ & $ap^*$ & $\delta_1(^\circ)$    \\
 \hline
 \hline                   &                      & 1 & 0.4559(52) &            &  0.1207(50)  & 137(3)   \\
                          &\raisebox{1.5ex}{CMF} & 2 & 0.6584(90) &            & 0.2686(57)   & 170(9)   
                          \\
                          &                      & 1 & 0.4869(35) & 0.4137(41) &  0.0729(61)  & 4.7(.3)   \\
   \raisebox{1.5ex}{$A_1$}&\raisebox{1.5ex}{MF1} & 2 & 0.5563(98)  & 0.494(11)  & 0.1543(91)  & 162(5)   
                          \\
                          &                      & 1 & 0.5660(42) & 0.4356(56) &  0.1000(62)  & 15.3(.4)   \\
                          &\raisebox{1.5ex}{MF2} & 2 & 0.642(11)  &0.533(13)   & 0.1838(97)    &160(6)     \\
 \hline
 \hline                   &                      & 1 &  0.4301(52)&            &  0.1331(42)  &128(3)   \\
                          &\raisebox{1.5ex}{CMF} & 2 & 0.637(16) &            & 0.2719(96)    & 165(15)   
                          \\
                          &                      & 1 & 0.4537(25) & 0.3737(31) &  0.0794(36)  & 4.4(.1)   \\
   \raisebox{1.5ex}{$A_2$}&\raisebox{1.5ex}{MF1} & 2 & 0.527(12)  & 0.461(14)  & 0.157(11)    & 159(6)  
                          \\
                          &                      & 1 & 0.5343(57) & 0.3925(80) &  0.0997(79)  & 12.9(.6)   \\
                          &\raisebox{1.5ex}{MF2} & 2 & 0.612(16)  &0.495(20)   & 0.182(14)    & 159(9)  \\

 \hline
 \hline                   &                      & 1 &  0.4037(68)&            &  0.1516(46)  & 70(6)   \\
                          &\raisebox{1.5ex}{CMF} & 2 & 0.4931(80) &            & 0.2081(48)   & 156(10)   
                          \\
                          &                      & 1 & 0.3638(13) & 0.3076(15) &  0.0761(16)  & 2.4(.4)   \\
   \raisebox{1.5ex}{$A_3$}&\raisebox{1.5ex}{MF1} & 2 & 0.474(23)  & 0.433(25)  & 0.171(16)    & 103(22)   
                          \\
                          &                      & 1 & 0.4330(25) & 0.3354(33) &  0.1013(27)  & 4(2)   \\
                          &\raisebox{1.5ex}{MF2} & 2 & 0.518(18)  &0.441(21)   & 0.176(13)    & 120(15)    \\
 \hline
 \hline                   &                      & 1 &  0.3844(79)&            &  0.1534(50)  & 67(7)   \\
                          &\raisebox{1.5ex}{CMF} & 2 & $0.4591(86)^*$ &            &              &         
                          \\
                          &                      & 1 & 0.3363(14) & 0.2743(17) &  0.0726(15)  & 2.4(.3)   \\
   \raisebox{1.5ex}{$A_4$}&\raisebox{1.5ex}{MF1} & 2 & 0.440(19)  & 0.396(22)  & 0.161(13)    & 116(16)   
                          \\
                          &                      & 1 & 0.4035(36) & 0.2959(50) &  0.0915(40)  & 6(2)   \\
                          &\raisebox{1.5ex}{MF2} & 2 & 0.490(22)  &0.407(27)   & 0.167(17)    & 128(17)    \\
 \hline
\end{tabular}
\end{center}
\caption{We give the P-wave scattering phase $\delta_1$
as extracted from the
energies of the ground state and the first excited state in the CMF, MF1
and MF2. We list the ensemble number, the energies $E_n$ and $E_{CM}$, 
the momentum $p^*$ and the scattering
phase $\delta_1$ (in units of degree). The single result marked by a star denotes that the
corresponding $E_{CM}$ is above the $4m_\pi$ threshold. We
therefore exclude that point from our calculations.}
\label{tab:rho_decay_results}
\end{table}

\bibliography{method_rho_decay}
\bibliographystyle{unsrt_nt}

\end{document}